\documentclass[11pt,fleqn]{article}
\usepackage{color}
\usepackage{pstricks}
\usepackage{amssymb}
\usepackage{graphicx} 
\usepackage{bbm}
\usepackage{longtable}
\usepackage[small,loose]{subfigure}
\usepackage[small]{caption2} 
\usepackage{fleqn} 
\usepackage{cite}
\usepackage{pifont}
\usepackage{paralist}

\usepackage{latexsym}
\usepackage[centertags]{amsmath}
\usepackage{amsfonts}
\usepackage{amscd}
\usepackage{epsfig}
\DeclareMathAlphabet{\scr}{U}{rsfs}{m}{n}

\advance \headheight by 3.0truept       
\setcaptionwidth{0.75\textwidth}
 
\DeclareMathOperator{\tr}{tr} 

\newcommand{\CenterObject}[1]{\ensuremath{\vcenter{\hbox{#1}}}}

\newcommand{\I}{\mathrm{i}}
\newcommand{\BmL}{\ensuremath{B\!-\!L} }

\newcommand{\E}[1]{\ensuremath{\mathrm{E}_{#1}}} 
\newcommand{\G}[1]{\ensuremath{\mathrm{G}_{#1}}}
\newcommand{\SO}[1]{\ensuremath{\mathrm{SO}(#1)}}
\newcommand{\SU}[1]{\ensuremath{\mathrm{SU}(#1)}}
\newcommand{\U}[1]{\ensuremath{\mathrm{U}(#1)}}
\newcommand{\Z}[1]{\ensuremath{\mathbbm{Z}_{#1}}} 

\allowdisplaybreaks

\begin{document}

\date{}
\title{\begin{flushright}
\ \\*[-80pt]
\begin{minipage}{0.3\linewidth}
\normalsize
CERN-PH-TH/2007-126\\
OHSTPY-HEP-T-07-003              \\
TUM-HEP-673/07 \\*[50pt]
\end{minipage}
\end{flushright}
{\bf\huge The Heterotic Road to the \\ MSSM with R parity}\\[0.8cm]}

\author{{\bf\normalsize
Oleg~Lebedev$^1$\!, Hans~Peter~Nilles$^2$\!, Stuart~Raby$^3$\!,
Sa\'ul~Ramos-S\'anchez$^2$\!,}\\{\bf\normalsize Michael~Ratz$^4$\!,
Patrick~K.~S.~Vaudrevange$^2$\!,
Ak{\i}n~Wingerter$^3$}\\[1cm]
{\it\normalsize
${}^1$ CERN, Theory Division, CH-1211 Geneva 23, Switzerland}\\[0.2cm]
{\it\normalsize
${}^2$ Physikalisches Institut der Universit\"at Bonn,}\\[-0.05cm]
{\it\normalsize Nussallee 12, 53115 Bonn,
Germany}\\[0.2cm]
{\it\normalsize
${}^3$ Department of Physics, The Ohio State University,}\\[-0.05cm]
{\it\normalsize
191 W.\ Woodruff Ave., Columbus, OH 43210, USA}\\[0.2cm]
{\it\normalsize
${}^4$ Physik Department T30, Technische Universit\"at M\"unchen,}\\[-0.05cm]
{\it\normalsize James-Franck-Strasse, 85748 Garching, Germany} }
\maketitle \thispagestyle{empty}

\begin{abstract}
{ In a previous paper, referred to as a ``Mini-Landscape" search, we
explored a ``fertile patch" of the heterotic landscape based on a
\Z6-II orbifold with SO(10) and \E6 local GUT structures.  In the
present paper we extend this analysis. We find many models with the
minimal supersymmetric standard model spectra and an exact R parity.
In all of these models, the vector-like exotics decouple along D
flat directions. We present two ``benchmark" models which satisfy
many of the constraints of a realistic supersymmetric model,
including non-trivial Yukawa matrices for 3 families of quarks and
leptons and Majorana neutrino masses for right-handed neutrinos with
non-trivial See-Saw masses for the 3 light neutrinos. In an appendix
we comment on the important issue of string selection rules and in
particular the so-called ``gamma-rule". }
\end{abstract}
\clearpage

\numberwithin{equation}{section}

\section{Introduction}

The string landscape denotes the space of string vacua
\cite{Susskind:2003kw,Schellekens:2006xz,Lust:2007kw}.  It is
believed that there are on the order of  $10^{500}$ or more possible
vacuum configurations. However, at the moment, only a paltry few
such vacua have properties looking anything like our own, i.e.\ with
3 large space dimensions, the standard model gauge interactions and
matter degrees of freedom, and a vanishingly small (in string units)
cosmological constant. Much effort has gone into exploring the
landscape in search of regions satisfying the latter requisite
feature, while only a few groups have attempted to find the rest. If
the string is to make contact with experiment, this situation must
be inverted.

Can string theory with $10^{500}$ vacua make any predictions
relevant for the LHC?  It has been suggested that by exploring the
\emph{entire} string landscape one might obtain statistical data
which could lead to probabilistic experimental statements
\cite{Douglas:2003um,Douglas:2004qg}.   Yet the clearest statement
to-date is that standard-like models are exceedingly rare. This has
been demonstrated in the context of orientifolds of Gepner models
\cite{Dijkstra:2004cc,Anastasopoulos:2006da} and in the context of
intersecting D--branes in an orientifold background
\cite{Gmeiner:2005vz,Douglas:2006xy,Douglas:2006za,Gmeiner:2007we}.
Nevertheless they may be more prevalent in the heterotic string
because of the simple fact that GUT groups, such as SO(10)
containing spinor representations, appear naturally. For
explorations of the string landscape in the context of the free
fermionic construction of the heterotic string, see
\cite{Faraggi:2004rq,Dienes:2006ut,Faraggi:2006bc,Dienes:2006ca,Dienes:2007ms,Coriano:2007ba}.
However, even within the context of the heterotic string, standard
model-like structure is not guaranteed. For example, it appears to
be very difficult \cite{Giedt:2000bi,Giedt:2005vx}, if not
impossible \cite{Ibanez:1992hc,Araki:2007ss}, to find standard
model-like structure in the heterotic string compactified on a
$\mathbb{Z}_3$ orbifold.  To summarize, standard-like models are
very rare; quite possibly a negligibly small set in the
\emph{entire} landscape. If so, what might we conclude from this
statistic?

We suggest the following alternate strategy for obtaining low energy predictions
from string theory.   One should introduce some \emph{priors} into ones
statistical analysis.   The first \emph{prior} is that the theory has only 3
large space dimensions.  The second is that the string vacuum includes the
standard model. Perhaps within this subset one may find some statistical
correlations which can be useful. Indeed, it is also important to verify that
the standard model actually sits in the string landscape.   Of course, in order
to make this analysis tractable, one may need to include additional
\emph{priors}.   In particular, one may require that the theory is
supersymmetric at the string scale and that below the string scale the spectrum
is that of the MSSM. Such theories typically have of order 100 moduli
(geometric or others). The Yukawa and gauge couplings of the theory will
generically be functions of these moduli.   In the supersymmetric limit of the
theory, one would hope to be able to tune the moduli in order to obtain
acceptable low energy physics. Of course, the problem of stabilizing moduli and
supersymmetry breaking must be addressed. However, it is clear that if one
\emph{cannot} find an MSSM-like model with this caveat, then this class of
theories can be ignored.\footnote{Note, a handful of heterotic string models
with MSSM-like structure have been discussed in the literature
\cite{Chaudhuri:1995ve,
Pokorski:1998hr,
Cleaver:1998sa,
Cleaver:1999cj,
Cleaver:1999mw,
Donagi:1999ez,
Donagi:2004su,
Donagi:2004ub,
Bouchard:2005ag,
Braun:2005nv,
Blumenhagen:2006ux,
Bouchard:2006dn,
Kim:2006hv, Kim:2006zw, Blumenhagen:2006wj, Kim:2007mt,
Cleaver:2007ek,Munoz:2007sf}.} Finally,  the cosmological constant
problem would still need to be addressed. But perhaps the only role
of the $10^{500}$ vacua is to resolve this problem.

In a previous paper, ``Mini-Landscape" [ML] \cite{Lebedev:2006kn} we advocated
this landscape philosophy.  The present paper extends the previous search and
also addresses some important phenomenological issues. We base our model scan on
the heterotic $\E8\times\E8$ string \cite{Gross:1984dd,Gross:1985fr}
compactified on an orbifold
\cite{Dixon:1985jw,Dixon:1986jc,Ibanez:1986tp,Ibanez:1987xa,Ibanez:1987sn,
Casas:1988hb,Casas:1987us}. Our study is motivated by recent work on an orbifold
GUT interpretation of heterotic string models
\cite{Kobayashi:2004ud,Forste:2004ie,Kobayashi:2004ya}. We focus on the \Z6-II
$ \equiv \mathbb{Z}_3 \times \mathbb{Z}_2$ orbifold, which is described in
detail in \cite{Kobayashi:2004ud,Kobayashi:2004ya,Buchmuller:2006ik}. The search
strategy is based on the concept of ``local GUTs''
\cite{Buchmuller:2004hv,Buchmuller:2005jr,Buchmuller:2005sh,Buchmuller:2006ik,Buchmuller:2007qf}
which inherits certain features of standard grand unification
\cite{Pati:1974yy,Georgi:1974sy,Fritzsch:1974nn,Georgi:1975qb}. Local GUTs  are
specific to certain points in the compact space, while the 4D gauge symmetry is
that of the SM. If matter fields are localized at such points, they form a
complete GUT representation. This applies, in particular, to a
$\boldsymbol{16}$--plet of a local SO(10), which comprises one generation of the
SM matter plus a right--handed neutrino \cite{Georgi:1975qb,Fritzsch:1974nn},
\begin{equation}
\boldsymbol{16}~=~ (\boldsymbol{3}, \boldsymbol{2})_{1/6} +
(\boldsymbol{\overline{3}}, \boldsymbol{1})_{-2/3}+
(\boldsymbol{\overline{3}}, \boldsymbol{1})_{1/3}+ (\boldsymbol{1},
\boldsymbol{2})_{-1/2} + (\boldsymbol{1}, \boldsymbol{1})_{1}+
(\boldsymbol{1}, \boldsymbol{1})_{0}
 \;,
\end{equation}
where representations with respect to
$\SU3_\mathrm{C}\times\SU2_\mathrm{L}$ are shown in parentheses and
the subscript denotes hypercharge (with electric charge given by $Q
=\mathsf{T}_{3\mathrm{L}} + Y$). On the other hand, bulk fields are
partially projected out and form incomplete GUT multiplets. This
offers an intuitive explanation for the observed multiplet structure
of the SM
\cite{Buchmuller:2004hv,Buchmuller:2005jr,Buchmuller:2005sh,Buchmuller:2006ik}.
This framework is consistent with MSSM gauge coupling unification as
long as the SM gauge group is embedded in a simple local GUT
$G_\mathrm{local}\supseteq \SU5$, which leads to the standard
hypercharge normalization.\footnote{Note even if one relaxes this
constraint as a prior it was shown that 90\% of the MSSM-like models
satisfying $\sin^2\theta_W = 3/8$ at the string scale necessarily
satisfy this constraint \cite{Raby:2007yc}. Also, the discrepancy
between the string scale, $\mathcal{O}(10^{17}\,\mathrm{GeV}$), and
the 4D GUT scale, $\mathcal{O} (10^{16}\,\mathrm{GeV}$), can in
principle be resolved by threshold corrections due to states near
the string scale.}

We find that the above search strategy, as opposed to  a random
scan, is successful and a considerable fraction of the models with
SO(10) and  \E6 local GUT structures pass our criteria. Out of about
$ 3\times10^4$ inequivalent models which involve 2 Wilson lines,
$\mathcal{O}(200)$ are phenomenologically attractive and can serve
as an ultraviolet completion of the MSSM.  In the present paper we
extend our previous analysis in several ways.
\begin{itemize}

\item In ML \cite{Lebedev:2006kn}, at the last step in our analysis of a theory,
we evaluated the effective mass operators, for the vector-like
exotics, up to order 8 in fields. If all the exotics
obtained mass, the model was retained. When calculating the rank of the mass
matrices, we  \emph{assumed} that requiring the singlet configuration to respect
supersymmetry would not change the result.

In this paper we explicitly demonstrate that the decoupling of the exotics is
consistent with supersymmetry.  We first find the $D = 0$ flat
directions.   If the exotics decouple along these directions, then
in particular models we check for $F = 0$.  Then complexified
gauge transformations allow us to satisfy $F=0$ and $D=0$ simultaneously.

\item  In ML \cite{Lebedev:2006kn}, we presented a model allowing for R parity.
However, we did not perform a systematic search for R parity invariant vacua.
Dangerous R parity violating dimension four operators can be forbidden by
\emph{family reflection symmetry} (FRS) [or \emph{matter parity}]
\cite{Dimopoulos:1981dw}, i.e.\ a discrete $\mathbb{Z}_2$ subgroup of $\U1_{B -
L}$ (baryon minus lepton number). For other approaches see
\cite{Gaillard:2004aa,Araki:2007ss}.

In this paper we evaluate $B - L$, searching for a ``suitable"
definition which has the accepted value on all standard model
particles and gives most standard model singlets a value satisfying
$3 (B - L) = 0 \mod 2$.  This condition preserves a
$\mathbb{Z}_2^{\cal M}$ subgroup of $B-L$ under which chiral matter
superfields are odd and Higgs superfields are even. Singlets with $3
(B - L) = 0\mod 2$ can obtain vacuum expectation values (VEVs) for
decoupling exotics, as well as giving effective quark and lepton
Yukawa couplings. Some of these singlets give Majorana masses to
right-handed neutrinos \cite{Buchmuller:2007zd} (for earlier work
see \cite{Giedt:2005vx}), preserving R-parity. Note,  if singlets
with $3 (B - L) = 1\mod 2$ obtain VEVs, R-parity is broken and
dimension 4 baryon/lepton number violating operators are typically
generated. In an Appendix, we also consider a possible
$\mathbb{Z}_N$ generalization of FRS.

\end{itemize}

The paper is organized as follows.  In Section~\ref{sec:strategy} we review the
search strategy defined in ML \cite{Lebedev:2006kn}. In Section
\ref{sec:decoupling} we present our results solely on the issue of decoupling of
vector-like exotics along $D$--flat directions. At this point we can compare our
results to other MSSM searches in different regions of the string landscape. We
show that we are extremely successful in finding models which have the
characteristics of the MSSM.  In the following sections we consider many of the
phenomenological issues one must face on the road to the MSSM. In particular, in
Section \ref{sec:rparity} we discuss the problem of obtaining one pair of light
Higgs doublets, a heavy top and then the additional constraint for a conserved
R-parity/{\em family reflection symmetry}.  In Section \ref{sec:yukawacouplings}
we discuss two models which satisfy the aforementioned constraints in detail. In
particular we consider the effective Yukawa couplings for quarks and leptons in
the limit that exotics decouple. We also study the See-Saw mechanism in these
examples.  Finally, in Section \ref{sec:conclusion} we summarize our results and
discuss some remaining issues.

\section{``Mini-Landscape" search strategy \cite{Lebedev:2006kn}: Local GUTs \label{sec:strategy}}

\begin{figure}[h]
\[
\CenterObject{\includegraphics{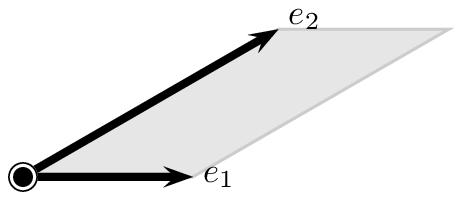}}
\times
\CenterObject{\includegraphics{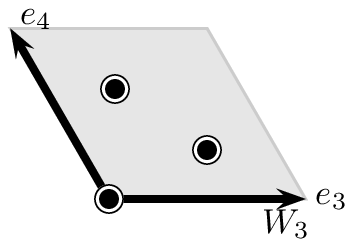}}
\times
\CenterObject{\includegraphics{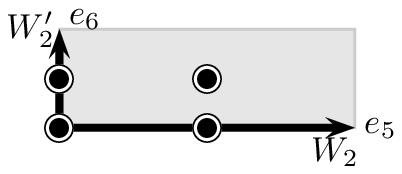}}
\]
\caption{$\G2\times\SU3\times\SO4$ torus lattice of a $\Z6$-II
orbifold. Possible Wilson lines are denoted by $W_3$, $W_2$ and
$W_2'$. The fixed points in the figure are those of the $T_1$
twisted sector.} \label{fig:T1}
\end{figure}

Our model search is carried out in the \Z6-II orbifold compactification of the
heterotic $\E8\times\E8$ string with the twist vector
\begin{equation}
 \vec{v}~=~\frac{1}{6} (1, 2, -3)
\label{eq:twist}
\end{equation}
acting on an \G2 $\times$ \SU3 $\times$ \SO4 torus (see Fig.~\ref{fig:T1}; for
details see \cite{Kobayashi:2004ya,Buchmuller:2006ik}).

It is well known that with a suitable choice of Wilson lines it is not difficult
to obtain the SM gauge group up to \U1 factors. The real challenge is to get the
correct matter spectrum and  the GUT hypercharge normalization.  To this end, we
base our strategy on the concept  of local GUTs. An  orbifold model is defined
by the orbifold twist, the torus lattice and the  gauge embedding of the
orbifold action, i.e.\ the gauge shift $V$ and the Wilson lines $W_n$.
We consider only the gauge shifts $V$ which allow for a local SO(10)
or \E6 structure, i.e. including ${\bf 16-}$ or ${\bf 27-}$plets in
the $T_1$ twisted sector. For the twist \eqref{eq:twist}, the SO(10)
shifts are given by \cite{Katsuki:1989cs}
\begin{eqnarray}
 V^{ \SO{10},1} &= &
 \left(\tfrac{1}{3},\,\tfrac{1}{2},\,\tfrac{1}{2},\,0,\,0,\,0,\,0,\,0\right)\,\left(\tfrac{1}{3},\,0,\,0,\,0,\,0,\,0,\,0,\,0\right)
 \;,
 \nonumber \\
 V^{ \SO{10},2 }& = &
 \left(\tfrac{1}{3},\,\tfrac{1}{3},\,\tfrac{1}{3},\,0,\,0,\,0,\,0,\,0\right)\,\left(\tfrac{1}{6},\,\tfrac{1}{6},\,0,\,0,\,0,\,0,\,0,\,0\right)
 \;, \label{eq:so10shifts}
\end{eqnarray}
while the \E6 shifts read
\begin{eqnarray}
 V^{\E6 , 1} & = &
 \left(\tfrac{1}{2},\,\tfrac{1}{3},\,\tfrac{1}{6},\,0,\,0,\,0,\,0,\,0\right)\,\left(0,\,0,\,0,\,0,\,0,\,0,\,0,\,0\right)\;,
 \nonumber \\
 V^{ \E6 ,2}& = &
 \left(\tfrac{2}{3},\,\tfrac{1}{3},\,\tfrac{1}{3},\,0,\,0,\,0,\,0,\,0\right)\,\left(\tfrac{1}{6},\,\tfrac{1}{6},\,0,\,0,\,0,\,0,\,0,\,0\right)
 \;.\label{eq:e6shifts}
\end{eqnarray}

These gauge shifts are such that the left--moving momenta $P$ (we use the
standard notation, for details see
e.g.~\cite{Forste:2004ie,Kobayashi:2004ya,Buchmuller:2006ik}) satisfying
\begin{equation}\label{eq:LocalGaugeSymmetry}
 P \cdot V ~=~ 0\mod 1 \;,\quad P^2~=~2\;,\quad P\in\Lambda_{\E8\times\E8}
\end{equation}
are roots of SO(10) or \E6 (up to extra group factors). In fact, this defines
the ``local" gauge symmetry in the $T_1$ sector, for states residing at the
origin in the \G2 and \SU3 tori and at the two fixed points in the \SO4 torus
which are unaffected by the $W_2$ Wilson line along the $e_6$ direction (see
Fig.~\ref{fig:T1}).\footnote{We assume that there are only two Wilson lines,
$W_3$ in the \SU3 torus and $W_2$ in the \SO4 torus.} The massless states of the
first twisted sector are guaranteed to contain $\boldsymbol{16}$--plets of
SO(10) at the fixed points with SO(10) symmetry or $\boldsymbol{27}$--plets of
\E6 at the fixed points with \E6 symmetry.  This is established by considering
the mass operator for left-movers,
\begin{equation}
\frac{1}{2} \ (P + V)^2 - 1 + \frac{1}{2} \sum_{i= 1}^3 |v_i| (1 - |v_i|)~=~0\;,
\end{equation}
with the shift vectors $V$ (Eqns.~\eqref{eq:so10shifts} and
\eqref{eq:e6shifts})  and the  twist vector $\vec{v}$ (Eq.~\eqref{eq:twist}).
For example, in the case of the two SO(10) shifts, the massless SO(10) spinor is
given by
\begin{equation}
 P~=~\frac{1}{2} (-1, -1,-1, \pm 1, \pm 1, \pm 1, \pm 1, \pm 1)\,(0^8)
\end{equation}
with an even number of minus signs.

Since these massless states are automatically invariant under the orbifold
action, they all survive in 4D and appear as complete GUT multiplets. In the
case of \SO{10}, that gives two complete SM generations, while in the case of
\E6 we have two $\boldsymbol{27}$s with
$\boldsymbol{27}=\boldsymbol{16}+\boldsymbol{10}+\boldsymbol{1}$ under \SO{10}.
It is thus necessary to decouple all (or part) of the $\boldsymbol{10}$s from
the low energy theory.  The third generation has to come from other twisted or
untwisted sectors.  The localized $\boldsymbol{16}$-- and
$\boldsymbol{27}$--plets are true GUT multiplets, whereas the third or
``bulk''   generation  only has the SM quantum numbers of an additional
$\boldsymbol{16}$--plet.

The Wilson lines are chosen such that the standard model gauge group is embedded
into the local GUT as
\begin{equation}
 G_\mathrm{SM}~ \subset~ \SU5 \subset \SO{10} ~\text{or}~ \E6 \;.
\end{equation}
Moreover, hypercharge is that of standard GUTs and thus consistent
with gauge coupling unification. The spectrum has certain features
of traditional 4D GUTs, e.g.\ matter fields form complete GUT
representations, yet there are important differences. In particular,
interactions generally break GUT relations since different local
GUTs are supported at different fixed points. Also, gauge coupling
unification is due to the fact that the 10D (not 4D) theory is
described by a single coupling.

Let us now recall the search strategy and results from the ``Mini-Landscape"
search \cite{Lebedev:2006kn}.  Consider, for example, models with the \SO{10}
local structure. For each of the \SO{10} shifts of Eq.~\eqref{eq:so10shifts}, we
follow the steps:
\begin{dingautolist}{"0C0}
 \item Generate Wilson lines $W_3$ and $W_2$.
 \item Identify ``inequivalent'' models.
 \item Select models with $G_\mathrm{SM} \subset \SU5 \subset \SO{10}$.
 \item Select models with three net
 $(\boldsymbol{3},\boldsymbol{2})$.
 \item Select models with non--anomalous $\U1_{Y} \subset \SU5$.
 \item Select models with net 3 SM families + Higgses + vector--like.
\end{dingautolist}
The results are presented in table \ref{tab:Summary}. The models
with the chiral MSSM matter content are listed in
\cite{WebTables:2006ml}.\footnote{In a recent paper
\cite{Raby:2007yc}, the two constraints,  \ding{"0C2} SM gauge group
$\subset$ SU(5) $\subset$ SO(10) ({\rm or}~\E6) and \ding{"0C4}
non--anomalous $\U1_{Y}\subset \SU5 $ were removed.   This search
has lead to about 10 times more models.   However the additional
constraint that $\sin^2\theta_W = 3/8$ reduced this number by 90\%
so that there were only a handful of additional models.   It
suggests that in order to find the MSSM, one may need to require
local GUTs.}

\begin{table*}[h!]
\centerline{
\begin{tabular}{l||l|l||l|l}
 criterion & $V^{\SO{10},1}$ & $V^{\SO{10},2}$ & $V^{\E6,1}$ & $V^{\E6,2}$\\
\hline
&&&&\\
 \ding{"0C1}  inequivalent models with 2 Wilson lines
  &$22,000$ & $7,800$  &$680$  &$1,700$ \\[0.2cm]
  \ding{"0C2} SM gauge group $\subset$ SU(5) $\subset$ SO(10)
  ({\rm or}~\E6)
  &3563 &1163 &27 &63\\[0.2cm]
  \ding{"0C3} 3 net $(\boldsymbol{3},\boldsymbol{2})$
  &1170 &492 &3 &32\\[0.2cm]
  \ding{"0C4} non--anomalous $\U1_{Y}\subset \SU5 $
  &528 &234 &3 &22\\[0.2cm]
  \ding{"0C5}  spectrum $=$ 3 generations $+$ vector-like
  &128 &90 &3 &2
  \\
\hline
\end{tabular}
}
\caption{Statistics of \Z6-II orbifolds based on the shifts
$V^{\SO{10},1},V^{\SO{10},2},V^{\E6,1},V^{\E6,2}$ with two Wilson
lines. \label{tab:Summary} }
\end{table*}

To show that the decoupling of exotics is consistent with string selection rules
is a technically involved and time consuming issue. We must select models in
which the mass matrices for the exotics have a maximal rank such that no exotic
states appear at low energies.  We consider superpotential couplings up to order
6 in SM singlets. In our previous analysis, ML,  we allowed any SM singlet to
obtain a non-vanishing VEV.  In the following section we refine our search and
demand that all singlet VEVs be along $D$--flat directions. This requires
solving the non-trivial $D$--flatness conditions.  In this analysis we focus on
the two SO(10) shifts.  Note, there are 218 models in this sector after step
\ding{"0C5}. In the following section we consider the decoupling of exotics.
We do this in two steps.  In the first step we construct the effective mass
operators for the exotics and check to see if the exotics decouple allowing
arbitrary singlet VEVs.   In the second step we only consider singlet VEVs along
$D$--flat directions.

\section{Decoupling exotics}
\label{sec:decoupling}

We evaluate all effective mass operators for the exotics $x_i$, $\overline{x}_j$
up to order 6 in SM singlet fields $\widetilde{s}_i$,
\begin{equation}
 W~\supset~x_i\,\overline{x}_j\,\langle\widetilde{s}_1\cdots\widetilde{s}_N\rangle
 \;.\label{eq:Wexotic}
\end{equation}
In general, $\widetilde{s}$ transform non-trivially under the extra
\U1s and hidden sector gauge groups. To construct the mass operators
\eqref{eq:Wexotic}, we find all monomials of the above form
consistent with string selection rules. These rules have been
discussed previously in the literature. They include space group and
R-charge selection rules, in addition to the standard field
theoretic requirement of gauge invariance. Complete details of these
string selection rules are given in Appendix A.  We should emphasize
here that in previous analyses a $\gamma$ selection rule has also
been enforced \cite{Casas:1991ac,Kobayashi:1994ks,Kobayashi:2004ya}.
We disagree with this additional $\gamma$ rule and in Appendix A we
give a general argument why this rule is {\em not} a selection
rule.\footnote{In fact, we have shown that all exotics decouple in
model A1 in Ref. \cite{Kobayashi:2004ya} if one eliminates the
$\gamma$ rule.}

We consider the 218 models remaining after step \ding{"0C5} from the
two SO(10) shifts (128 from $V^{\SO{10},1}$ and 90 from
$V^{\SO{10},2}$), see Table \ref{tab:Summary}. If in a particular
model all exotics decouple to order 6 in the product of
$\widetilde{s}$ fields, assuming arbitrary $\widetilde{s}$ VEVs
\footnote{Note that giving VEVs to the $\widetilde{s}$ fields can
often be interpreted as blowing up the orbifold singularities (for
recent developments in this direction see
\cite{Lust:2006zh,Honecker:2006qz,Nibbelink:2007rd,Nibbelink:2007pn}).},
we retain the model. The number of models satisfying decoupling at
this step is 191 (106 from $V^{\SO{10},1}$ and 85 from
$V^{\SO{10},2}$). We now determine $D$--flat directions for all
$\widetilde{s}$ fields. Our procedure for determining $D$--flat
directions is described in Appendix B. We then retain the subset of
the 191 models for which the exotics decouple along $D$--flat
directions to order 6 in the $\widetilde{s}$ fields.  We find 190
models remaining. Clearly, $D$--flatness does not impose an
important constraint. Thus we are successful in
$190\big/3\cdot10^{4}$ or 0.6\,\% of the cases.

The results of our search may now be compared to many other searches
in the literature.  We have 218 models with the SM gauge group,  3
families and only vector-like exotics from our two SO(10) shifts.
Out of these we find 190 for which all exotics decouple along
$D$--flat directions.  In certain types of intersecting D--brane
models, it was found that the probability of obtaining the SM gauge
group and three generations of quarks and leptons, while allowing
for chiral exotics, is less than $10^{-9}$
\cite{Gmeiner:2005vz,Douglas:2006xy}. The criterion which comes
closest to the requirements imposed in
\cite{Gmeiner:2005vz,Douglas:2006xy} is \ding{"0C3}.  We find that
within our sample the corresponding probability is 6\,\%. In
\cite{Dijkstra:2004cc,Anastasopoulos:2006da}, orientifolds of Gepner
models were scanned for chiral MSSM matter spectra, and it was found
that the fraction of such models is $4 \times 10^{-14}$. These
constructions contain the MSSM matter spectrum plus, in general,
vector-like exotics.   This is most similar to step \ding{"0C5} in
our analysis where we find 218 models out of a total of ~ $3 \times
10^4$ or 0.7\,\%. In comparison, approximately 0.6\,\% of our models
have the MSSM spectrum at low energies with all vector-like exotics
decoupling (with exotic mass terms evaluated to order
$\widetilde{s}^6$) along $D$--flat directions. Note also that, in
all of our models, hypercharge is normalized as in standard GUTs and
thus consistent with gauge coupling unification.

\section{Road to the MSSM}
\label{sec:road}

In this section we consider other phenomenological hurdles which
must be overcome in order to reach the MSSM.   These hurdles
include finding supersymmetric minima with proton stability, an
exactly conserved R-parity, a $\mu$ term for the light Higgs
doublets of order the weak scale, a top quark Yukawa coupling of
order 1, gauge coupling unification, and more.

\subsection{Constraints}

\subsubsection*{R-parity conservation \label{sec:rparity}}

One of the most formidable obstacles in string constructions is
obtaining a conserved R-parity. In this paper we propose one
possible route, i.e.\ obtaining a ``family reflection symmetry" or
``matter parity".  In this regard, we evaluate $B - L$, searching
for a ``suitable" definition which has the accepted value on all
standard model particles, is vector-like on all exotics and produces
a number of SM singlets with even and zero $3 (B-L)$ charge.

Giving such singlets VEVs preserves a $\mathbb{Z}_2^{\cal M}$ subgroup of $B-L$,
denoted \emph{family reflection symmetry} or \emph{matter parity}, under which
chiral matter superfields are odd and Higgs superfields are even. We find that
the exotics can be decoupled and the right--handed neutrinos can be given
Majorana masses consistent with this symmetry. In Appendix C, we show that it is
possible to allow any $\widetilde{s}$ field to obtain a VEV as long as it has
$B-L$ eigenvalue $f =0,\pm 2/(2 \mathbb{Z} + 1)$.  This will leave invariant
$\mathbb{Z}_2^{\cal M}$.

To apply the above strategy, we must first give a ``suitable" definition of $B -
L$.  A possible algorithm to identify the corresponding generators is discussed
in Appendix D. Upon defining $B - L$, we must verify $D$--flatness for the
subset of SM singlets with $B - L$ charges $f =0,\pm 2/(2 \mathbb{Z} + 1)$ and
check that all exotics decouple. This is a tedious task, requiring much computer
time.   In order to minimize the amount of time, we focus our attention on a
subset of the 190 models which have renormalizable top quark Yukawa couplings.

\subsubsection*{Light Higgs doublets}

The Higgs doublets of the MSSM are vector-like and generically in our analysis
all the Higgs doublets decouple. Retaining one pair of light Higgs doublets in
the MSSM is the $\mu$ problem, and we must now face this issue.  We look for
vacuum configurations in which the $\mu$ term vanishes to a certain order in the
$\widetilde{s}$ fields. At the same time we require that all the exotics
decouple. Of course, it would be nice to have a symmetry argument for a small
$\mu$.

\subsubsection*{Order one top quark Yukawa coupling}

The top quark Yukawa coupling is necessarily of order 1.   Hence it is natural
(although perhaps not absolutely necessary) to require that for the top quark we
have a renormalizable $\mathcal{O}(1)$ Yukawa coupling $(\boldsymbol{3},
\boldsymbol{2})_{1/6}\,  (\boldsymbol{\bar 3}, \boldsymbol{1})_{-2/3} \,
(\boldsymbol{1}, \boldsymbol{2})_{1/2}$, i.e.\ one of the following types
\begin{equation}
 U\,U\,U\;,\quad U\,T\,T\;,\quad T\,T\,T \;,
\end{equation}
where $U$ and $T$ denote generic untwisted and twisted fields,
respectively. The $U\,U\,U$ coupling is given by the gauge coupling,
$U\,T\,T$ is a local coupling and thus is unsuppressed, while the
$T\,T\,T$ coupling is significant only when the twisted fields are
localized at the same fixed point. We discard models in which the
above couplings are absent or suppressed.  In ML we required that
the top quark have a Yukawa coupling at tree level, i.e.\ cubic
order in the fields, in addition to decoupling of all exotics,
albeit assuming arbitrary VEVs for the SM singlets.   Of the 190
models which decouple along $D$--flat directions we have 105 (85)
coming from the first (second) SO(10) shift. Out of these there are
55 (32) with ``heavy top" and 50 (53) with ``no heavy top." We thus
find 87 models which decouple along $D$--flat directions and have a
``heavy top."   Note, this is just one less than discussed in ML at
step \ding{"0C7}.

\subsubsection*{R-parity invariant models with cubic top Yukawa coupling}

We find a ``suitable" definition of $B - L$  for  34 of the 55 (5 of
the 32) models of the first (second) SO(10) shifts.   Note however
that for each case there are several possible inequivalent choices.
This is because of two ambiguities which need to be resolved.
\begin{enumerate}
\item In many cases there are vector-like exotics with SM gauge charges
identical to those of quarks, leptons and Higgs doublets.  Thus there are
different ways to choose which of these states have standard $B - L$
charges. Each choice can lead to a different definition of $B - L$.
\item  For each choice of SM particles above,  there may be more
than one $B - L$ definition.  In some cases there are continuous families of
solutions.
\end{enumerate}
Including all of these possibilities we find  3447 (144) suitable $B
- L$ generators from the first (second) SO(10) shifts, which also
lead to the presence of SM singlets with charges $B - L =0,\pm 2,
\pm2/3,\pm 2/5,\ldots, \pm 6/7$.  We find, however, that these lead
to 85 (8) inequivalent models. Requiring the absence of extra
unbroken \U1s reduces this set to 42 (0) acceptable models. Finally,
demanding that all exotics decouple along $D$--flat directions leads
to 15 (0) acceptable solutions with an exact low energy R-parity.
This result is specific to our $(B-L)$-based strategy and we expect,
in general, more acceptable models to exist.

\subsection{Approaching the MSSM}

Further issues to be addressed are as follows.
\begin{enumerate}
\item We must check that quarks and leptons obtain non-trivial masses.  For
neutrinos, this includes an analysis of Majorana masses and the See-Saw
mechanism.

\item We must also consider dimension 5 baryon and lepton number violating
operators \cite{Weinberg:1981wj,Sakai:1981pk,Dimopoulos:1981dw}.
These operators are not forbidden by R-parity and are typically
generated. Their coefficients must necessarily be suppressed in
order to be consistent with proton decay experiments
\cite{Hinchliffe:1992ad,Dermisek:2000hr}.

\item Precision gauge coupling unification should be addressed
\cite{Ibanez:1992hc,Hebecker:2004ce,Kobayashi:2004ya}. This includes a
calculation of the string threshold corrections
\cite{Dixon:1990pc,Mayr:1993mq,Stieberger:1998yi}.

\item  Finally,  $F=0$ has to be verified.  This constraint guarantees that our
vacua are indeed supersymmetric. In general, $F = 0$ solutions exist. Some of
them can be found numerically by truncating the superpotential and solving
polynomial equations. Once they are found, $F=0$ and $D=0$ can be satisfied
simultaneously using complexified gauge transformations \cite{Ovrut:1981wa} (for
a detailed discussion see \cite{Buchmuller:2006ik}).
\end{enumerate}

All of these checks are clearly time consuming and we have not
performed an inclusive analysis.  We have however found many vacua
with R-parity. In the next section we discuss our results for two
particular examples. In these examples we have demanded that:
\begin{itemize}
\item all exotics are massive,

\item there is one pair of massless Higgses,

\item the mass matrix for the right-handed neutrinos has full rank,

\item no extra U(1) factors remain,

\item hidden sector gaugino condensation is possible,

\item R-parity is unbroken.
\end{itemize}

\section{Two ``Benchmark" models with R-parity \label{sec:yukawacouplings}}

We now discuss two particular ``benchmark" models.   In Model 1 we
also consider two different vacuum configurations and show how the
phenomenology depends on the different choices of vacua.

\subsection{Model 1}

The model is defined by the shifts and Wilson lines given in
Appendix E. The gauge group after compactification is
\begin{equation}
 G_\mathrm{orbifold}~=~\SU3\times\SU2\times[\SU4\times\SU2']\times\U1^9\;.
\end{equation}

The resulting massless spectrum includes three SM generations plus
vector-like exotics with respect to the SM gauge group.

The  model  allows us  to define a  ``suitable" $B-L$ generator,
\begin{equation}
\mathsf{t}_{B-L} ~=~ \left(0 , 0 , 0 , 0 , 0 , -\frac{2}{3} ,
-\frac{2}{3} , -\frac{2}{3}\right)\, \left( 0 , 0 , 0 , 0 , 0 , 2 ,
0 , 0\right)\;.
\end{equation}
with two essential properties (cf.\ Table~\ref{tab:naming3}):
\begin{itemize}
\item the spectrum includes 3 generations of quarks and leptons plus vector-like exotics
with respect to $G_\mathrm{SM}\times\U1_{\BmL}\;,$ and
\item there are SM singlets with \BmL\ charge $\pm2$.
\end{itemize}

\begin{table}[h]
\centerline{%
\begin{tabular}{|c|l|l|c|c|l|l|}
\hline
\# & irrep & label & & \# & irrep & label\\
\hline
 3 &
$\left(\boldsymbol{3},\boldsymbol{2};\boldsymbol{1},\boldsymbol{1}\right)_{(1/6,1/3)}$
 & $q_i$
 & &
 3 &
$\left(\overline{\boldsymbol{3}},\boldsymbol{1};\boldsymbol{1},\boldsymbol{1}\right)_{(-2/3,-1/3)}$
 & $\bar u_i$
 \\
 3 &
$\left(\boldsymbol{1},\boldsymbol{1};\boldsymbol{1},\boldsymbol{1}\right)_{(1,1)}$
 & $\bar e_i$
 & &
 8 &
$\left(\boldsymbol{1},\boldsymbol{2};\boldsymbol{1},\boldsymbol{1}\right)_{(0,*)}$
 & $m_i$
 \\
 4 &
$\left(\overline{\boldsymbol{3}},\boldsymbol{1};\boldsymbol{1},\boldsymbol{1}\right)_{(1/3,-1/3)}$
 & $\bar d_i$
 & &
 1 &
$\left(\boldsymbol{3},\boldsymbol{1};\boldsymbol{1},\boldsymbol{1}\right)_{(-1/3,1/3)}$
 & $d_i$
 \\
 4 &
$\left(\boldsymbol{1},\boldsymbol{2};\boldsymbol{1},\boldsymbol{1}\right)_{(-1/2,-1)}$
 & $\ell_i$
 & &
 1 &
$\left(\boldsymbol{1},\boldsymbol{2};\boldsymbol{1},\boldsymbol{1}\right)_{(1/2,1)}$
 & $\bar \ell_i$
 \\
 1 &
$\left(\boldsymbol{1},\boldsymbol{2};\boldsymbol{1},\boldsymbol{1}\right)_{(-1/2,0)}$
 & $\phi_i$
 & &
 1 &
$\left(\boldsymbol{1},\boldsymbol{2};\boldsymbol{1},\boldsymbol{1}\right)_{(1/2,0)}$
 & $\bar \phi_i$
 \\
 6 &
$\left(\overline{\boldsymbol{3}},\boldsymbol{1};\boldsymbol{1},\boldsymbol{1}\right)_{(1/3,2/3)}$
 & $\bar\delta_i$
 & &
 6 &
$\left(\boldsymbol{3},\boldsymbol{1};\boldsymbol{1},\boldsymbol{1}\right)_{(-1/3,-2/3)}$
 & $\delta_i$
 \\
 14 &
$\left(\boldsymbol{1},\boldsymbol{1};\boldsymbol{1},\boldsymbol{1}\right)_{(1/2,*)}$
 & $s^+_i$
 & &
 14 &
$\left(\boldsymbol{1},\boldsymbol{1};\boldsymbol{1},\boldsymbol{1}\right)_{(-1/2,*)}$
 & $s^-_i$
 \\
 16 &
$\left(\boldsymbol{1},\boldsymbol{1};\boldsymbol{1},\boldsymbol{1}\right)_{(0,1)}$
 & $\bar n_i$
 & &
 13 &
$\left(\boldsymbol{1},\boldsymbol{1};\boldsymbol{1},\boldsymbol{1}\right)_{(0,-1)}$
 & $n_i$
 \\
 5 &
$\left(\boldsymbol{1},\boldsymbol{1};\boldsymbol{1},\boldsymbol{2}\right)_{(0,1)}$
 & $\bar \eta_i$
 & &
 5 &
$\left(\boldsymbol{1},\boldsymbol{1};\boldsymbol{1},\boldsymbol{2}\right)_{(0,-1)}$
 & $\eta_i$
 \\
 10 &
$\left(\boldsymbol{1},\boldsymbol{1};\boldsymbol{1},\boldsymbol{2}\right)_{(0,0)}$
 & $h_i$
 & &
 2 &
$\left(\boldsymbol{1},\boldsymbol{2};\boldsymbol{1},\boldsymbol{2}\right)_{(0,0)}$
 & $y_i$
 \\
 6 &
$\left(\boldsymbol{1},\boldsymbol{1};\boldsymbol{4},\boldsymbol{1}\right)_{(0,*)}$
 & $f_i$
 & &
 6 &
$\left(\boldsymbol{1},\boldsymbol{1};\overline{\boldsymbol{4}},\boldsymbol{1}\right)_{(0,*)}$
 & $\bar f_i$
 \\
 2 &
$\left(\boldsymbol{1},\boldsymbol{1};\boldsymbol{4},\boldsymbol{1}\right)_{(-1/2,-1)}$
 & $f_i^-$
 & &
 2 &
$\left(\boldsymbol{1},\boldsymbol{1};\overline{\boldsymbol{4}},\boldsymbol{1}\right)_{(1/2,1)}$
 & $\bar f_i^+$
 \\
 4 &
$\left(\boldsymbol{1},\boldsymbol{1};\boldsymbol{1},\boldsymbol{1}\right)_{(0,\pm2)}$
 & $\chi_i$
 & &
 32 &
$\left(\boldsymbol{1},\boldsymbol{1};\boldsymbol{1},\boldsymbol{1}\right)_{(0,0)}$
 & $s^0_i$
 \\
 2 &
$\left(\overline{\boldsymbol{3}},\boldsymbol{1};\boldsymbol{1},\boldsymbol{1}\right)_{(-1/6,2/3)}$
 & $\bar v_i$
 & &
 2 &
$\left(\boldsymbol{3},\boldsymbol{1};\boldsymbol{1},\boldsymbol{1}\right)_{(1/6,-2/3)}$
 & $v_i$
 \\
\hline
\end{tabular}
}
\caption{Spectrum. The quantum numbers under
$\SU3\times\SU2\times[\SU4\times\SU2']$ are shown in boldface; hypercharge and
\BmL\ charge appear as subscripts.  Note that the states $s_i^\pm$, $f_i$, $\bar
f_i$ and $m_i$ have different $B-L$ charges for different $i$, which we do not
explicitly list.}
\label{tab:naming3}
\end{table}

In the following discussion we consider two different vacuum
configurations to illustrate the dependence on the particular vacuum
class, i.e. the set of SM singlets with non-zero supersymmetric
VEVs.

\subsubsection{Model 1, vacuum configuration A}

Consider a vacuum configuration where the fields
\begin{eqnarray}
\{\widetilde{s}_i\} = & \{\chi_{1}, \chi_{2}, \chi_{3}, \chi_{4},
h_{1}, h_{2}, h_{3}, h_{4}, h_{5}, h_{6}, h_{9}, h_{10}, s^0_{1},
s^0_{4}, s^0_{5}, s^0_{6}, s^0_{9}, s^0_{11},
s^0_{13}, s^0_{15}, & \nonumber\\
 & \quad s^0_{16}, s^0_{17}, s^0_{18}, s^0_{20},
s^0_{21}, s^0_{22}, s^0_{23}, s^0_{25}, s^0_{26}, s^0_{27},
s^0_{30}, s^0_{31} \} &  \label{eq:vac1a}
\end{eqnarray}
develop a VEV while the expectation values of all other fields
vanish. In this vacuum configuration we set 14 of the original 46 SM
and hidden $\SU4$ singlets to zero.  The emerging effective theory
has the following properties:
\begin{enumerate}
\item the unbroken gauge symmetries are
\begin{equation}
 G_\mathrm{SM}\times G_\mathrm{hid}\;,
\end{equation}
where $G_\mathrm{hid}=\SU4$.
\item since \BmL\ is broken by two units, there is an effective
matter parity $\mathbb{Z}_2^{\cal M}$.
\item there is only one pair of Higgs candidates, $\phi_1$ and $\bar\phi_1$;
the $\mu$-term
\begin{equation}
 \mu~=~\left.\frac{\partial^2 W}{\partial\phi_1\,\partial\bar \phi_1}
 \right|_{\phi_1=\overline{\phi}_1=0}
\end{equation}
vanishes up to order $\widetilde{s}^6$, at which we work. That is,
there is one pair of massless Higgs doublets.
\item we check that the solution satisfies $F=0$ for all fields (cf.\ {\bf F-flatness} below)
and in addition, switching on $\{\widetilde{s}_i\}$-fields allows us
to cancel the FI term without inducing $D$-terms (cf.\
Appendix~\ref{sec:Dflat}).
\item all vector-like exotics decouple (cf.\ Appendix~\ref{sec:exoticsMasses1}).
\item neutrino masses are suppressed via the see-saw mechanism.
\end{enumerate}

That is, we have obtained a supersymmetric vacuum with the precise matter
content of the MSSM \emph{with an exact R parity}. This has to be contrasted to
\cite{Buchmuller:2006ik} where R parity was approximate, and to
\cite{Bouchard:2006dn} where R parity exists only at the classical level, and
where $Y_e$ and $Y_d$ vanish at the same level. Our model also does not suffer
from the problem encountered in \cite{Kim:2007mt}, where it was found that you
can either decouple all exotics or have $R$-parity but never both.

\subsubsection*{Charged fermion Yukawa matrices}

The charged fermion Yukawa matrices are\footnote{Quark and lepton
doublets multiply the Yukawa matrices on the left.}
\begin{eqnarray}
Y_u&=& \left(
\begin{array}{ccc}
\widetilde{s}^5 & \widetilde{s}^5 &  \widetilde{s}^5 \\
\widetilde{s}^5 & \widetilde{s}^5 &  \widetilde{s}^6 \\
\widetilde{s}^6 & \widetilde{s}^6 &  1
\end{array}
\right)\;,\quad Y_d~=~ \left(
\begin{array}{ccc}
 0 & \widetilde{s}^5 & 0 \\
 \widetilde{s}^5 & 0 & 0 \\
 0 & \widetilde{s}^6 & 0
\end{array}
\right)\;,\quad Y_e~=~ \left(
\begin{array}{ccc}
 0 & \widetilde{s}^5 & \widetilde{s}^6 \\
 \widetilde{s}^5  & 0 & 0 \\
 \widetilde{s}^6 & \widetilde{s}^6 & 0
\end{array}
\right)\;.\nonumber\\
& &
\end{eqnarray}
Here, $\widetilde{s}^n$ for $n$ a non-negative integer, represents
the the smallest value of $n$ for which this term appears in the
matrix; thus giving the dominant contribution. Each term is in fact
a sum of monomials containing several different SM and hidden $\SU4$
singlets.  The up-type quark Yukawa matrix is given directly in
terms of the coupling of the up-type Higgs to the three $q$ and
$\bar u$ fields. The down-type quark and charged lepton Yukawa
matrices are obtained by integrating out a pair of vector-like $d$-
and $\bar d$-quarks and $\ell$- and $\bar\ell$-fields, respectively.
We find that the up and charged lepton Yukawa matrices have rank 3,
while the down quark Yukawa matrix has only rank 2, at this order in
$\tilde s$ singlets. However, we have checked that at order 8 in
$\widetilde{s}$ fields $Y_d$ has rank 3.

\subsubsection*{Neutrino matrices}\label{sec:NeutrinoMasses1}

In our vacua $\SU2'$ is broken such that the $\SU2'$ doublets $\
\eta_i$ and $\bar\eta_i$ correspond to SM singlets with
$q_{\BmL}=\pm1$,
\begin{subequations}
\begin{equation}
 \bar\eta_1
 ~=~
 \left(\begin{array}{c}
 \bar n_{17}\\
 \bar n_{18}\end{array}\right)\;,\dots
 \bar\eta_5
 ~=~
 \left(\begin{array}{c}
 \bar n_{25}\\
 \bar n_{26}\end{array}\right)
\end{equation}
and
\begin{equation}
 \eta_1
 ~=~
 \left(\begin{array}{c}
  n_{14}\\
  n_{15}\end{array}\right)\;,\dots
 \eta_5
 ~=~
 \left(\begin{array}{c}
  n_{22}\\
  n_{23}\end{array}\right)
 \;.\nonumber\\
 \end{equation}
\end{subequations}

The dimensions of the ``right-handed" neutrino mass matrices are
\begin{eqnarray}
\mathcal{M}_{nn} & = & 23 \times 23 \nonumber\;,  \\
 \mathcal{M}_{n\bar n}  & = & 23 \times 26\;, \nonumber \\
\mathcal{M}_{\bar n\bar n}  & = &   26 \times 26\;,\label{eq:Mneutrino}
\end{eqnarray}
with the complete $\bar\nu-\bar\nu$ mass matrix given by
\begin{equation}
 \mathcal{M}_{\bar\nu\bar\nu}~=~
 \left(\begin{array}{cc}\mathcal{M}_{nn} & \mathcal{M}_{n\bar n}\\
 \mathcal{M}_{n\bar n}^T & \mathcal{M}_{\bar n\bar n}\end{array}\right)
 \;.
\end{equation}
We have checked that it indeed has full rank.  For more details, see Webpage
\cite{WebTables:2006ml}.

The Dirac neutrino Yukawa couplings have dimensions
\begin{eqnarray}
 Y_n
 & = & 4 \times 23\;,  \nonumber\\
 Y_{\bar n}
 & = &  4 \times 26\;.\label{eq:Yn}
\end{eqnarray}
The effective light neutrino mass operator emerges from
\begin{equation}
 \kappa~=~Y_\nu\,\mathcal{M}_{\bar\nu\bar\nu}^{-1}\,Y_\nu^{T}\;,
\end{equation}
where $Y_\nu~=~(Y_n,Y_{\bar n})$, by integrating out the pair of
heavy leptons $\bar \ell_1$ and $\ell'$ where $\ell'$ is a linear
combination of the $\ell_i$.   We have checked that the light
neutrinos all obtain a small mass.  The large dimension of the
matrices effectively reduces the See-Saw scale
\cite{Buchmuller:2007zd}. Also, neutrino phenomenology works
differently in the presence of many `right-handed' neutrinos
\cite{Eisele:2007ws,Ellis:2007wz}.

\subsubsection*{Dimension 5 baryon and lepton number violating operators}

We further analyzed the question of dimension 5 proton decay
operators. We find that both $q\,q\,q\,\ell$ and $\bar u\,\bar
u\,\bar d\,\bar e$ appear at order $\widetilde{s}^6$. They are also
generated by integrating out the heavy exotics.  For example, the
following couplings exist
\begin{equation}
  q_1 \,\ell_1\,\bar\delta_4\,,\; q_1 \,\ell_1\,\bar\delta_5\,,\; q_2
\,\ell_2\,\bar\delta_4\,,\;q_2
  \,\ell_2\,\bar\delta_5\,,\;
  q_1 \,q_1\,\delta_4\,,\;  q_1 \,q_1\,\delta_5\,,\;q_2
\,q_2\,\delta_4\,,\;q_2 \,q_2\,\delta_5\;.
\label{eq:trilinear_D5_operators1}
\end{equation}
Hence integrating out the states  $\bar\delta_i, \ \delta_i$
produces dangerous dimension 5 operators. These must be sufficiently
suppressed to be consistent with present bounds on proton decay
\cite{Hinchliffe:1992ad,Dermisek:2000hr}. We have verified that, for
some particular $\widetilde s$ VEVs, it is possible to suppress the
$q\,q\,q\,\ell$ operators induced by the trilinear
couplings~\eqref{eq:trilinear_D5_operators1}. However, higher order
couplings also introduce baryon and lepton number violating
operators. We have not been able to identify a suppression mechanism
for such operators yet.

\subsubsection*{$\boldsymbol{\mu}$-term and Minkowski space}

Since our singlet configuration satisfies $F=D=0$, the vacuum energy
is zero in the global SUSY limit. In supergravity, one should
include non--perturbative moduli potentials which would be
responsible for spontaneous SUSY breaking. In fact, in the context
of gaugino condensation
\cite{Ferrara:1982qs,Nilles:1982ik,Derendinger:1985kk,Dine:1985rz},
the SU(4) subgroup of the second \E8 gives rise to TeV soft masses,
which is a common feature of our ``fertile" patch of the landscape
\cite{Lebedev:2006tr}.  A specific realization of SUSY breaking via
gaugino condensation in heterotic string compactifications is given
by K\"{a}hler stabilization
\cite{Binetruy:1996xj,Casas:1996zi,Gaillard:2007jr}. Requiring a
Minkowski vacuum puts a constraint on the total superpotential which
includes contributions from charged matter and moduli.

An interesting feature of our singlet configuration  is that the charged matter
superpotential vanishes at order 6 in singlet fields,
\begin{equation}
 \left\langle  W(\widetilde{s} ) \right\rangle~=~0 \;.
\end{equation}
In fact the superpotential is a polynomial in $\widetilde{s}$ fields
and in this vacuum configuration each monomial term in $W$ vanishes
independently. Therefore, the total superpotential is given solely
by its non--perturbative part. This is expected to be very small and
thus a small gravitino mass and a small cosmological constant can in
principle be achieved.

In this model we also find  an intriguing correlation between the
$\mu$--term and $ W(\widetilde{s})$. Since the Higgs doublets are
untwisted and the combination $\phi_1\,\overline{\phi}_1$ has vacuum
quantum numbers, requiring that each monomial appearing in $\mu$
vanishes also implies $\langle W(\widetilde{s} ) \rangle
=0$.\footnote{This applies to the untwisted Higgs pairs in many
models of our MiniLandscape, for instance also to the model
presented in \cite{Buchmuller:2005jr,Buchmuller:2006ik}.} This means
that the $\mu$-term is of the order of the expectation value of $W$,
i.e.\ the gravitino mass.

\subsubsection*{$\boldsymbol{F}$-flatness \label{sec:Fflat}}

We analyze the $F$-terms in the configuration defined by
\eqref{eq:vac1a}. The only non-vanishing $F$-terms are
\begin{equation}
 F_i~=~\frac{\partial W}{\partial s^0_i}
 \quad\text{where}\:
 s^0_i\in\{s^0_3,s^0_7,s^0_{10},s^0_{14},s^0_{19}\}\;.
\end{equation}
They read
\begin{subequations}
\begin{eqnarray}
\scriptstyle F_{3} & \scriptstyle  = & \scriptstyle  \alpha_{27}\,s^0_{26} s^0_{4} +\alpha_{28}\,s^0_{15}
(s^0_{26})^2 s^0_{4} +\alpha_{29}\,s^0_{16} (s^0_{26})^2 s^0_{4}\;,\label{eq:F3}\\
\scriptstyle F_{7} & \scriptstyle  = & \scriptstyle  \alpha_{41}\,h_{1} h_{10} s^0_{30} s^0_{4}
+\alpha_{44}\,h_{2} h_{9} s^0_{30} s^0_{4} +\alpha_{46}\,s^0_{17}
s^0_{25} s^0_{30} s^0_{4} +\alpha_{47}\,s^0_{1} s^0_{18} s^0_{26}
s^0_{30} s^0_{4} +\alpha_{50}\,s^0_{18} s^0_{27} s^0_{30} s^0_{4}
+\alpha_{55}\,s^0_{15} (s^0_{30})^2 s^0_{4}
\nonumber\\
\scriptstyle & \scriptstyle  & \scriptstyle  {} +\alpha_{56}\,s^0_{16} (s^0_{30})^2 s^0_{4}
+\alpha_{57}\,s^0_{20} s^0_{30} s^0_{31} s^0_{4}
+\alpha_{58}\,s^0_{21} s^0_{30} s^0_{31} s^0_{4}\;,\\
\scriptstyle F_{10} & \scriptstyle  = & \scriptstyle  \alpha_{1}\,h_{1} h_{10} s^0_{13} s^0_{30}
+\alpha_{4}\,h_{2} h_{9} s^0_{13} s^0_{30} +\alpha_{12}\,s^0_{13}
s^0_{17} s^0_{25} s^0_{30} +\alpha_{13}\,s^0_{1} s^0_{13} s^0_{18}
s^0_{26} s^0_{30} +\alpha_{16}\,s^0_{13} s^0_{18} s^0_{27} s^0_{30}
\nonumber\\
\scriptstyle & \scriptstyle  & \scriptstyle  {} +\alpha_{21}\,s^0_{13} s^0_{15} (s^0_{30})^2
+\alpha_{22}\,s^0_{13} s^0_{16} (s^0_{30})^2 +\alpha_{23}\,s^0_{13}
s^0_{20} s^0_{30} s^0_{31} +\alpha_{24}\,s^0_{13} s^0_{21} s^0_{30}
s^0_{31} +\alpha_{81}\,s^0_{26} s^0_{9}
\nonumber\\
\scriptstyle & \scriptstyle  & \scriptstyle  {} +\alpha_{82}\,s^0_{15} (s^0_{26})^2 s^0_{9}
+\alpha_{83}\,s^0_{16} (s^0_{26})^2 s^0_{9}\;,\\
\scriptstyle F_{14} & \scriptstyle  = & \scriptstyle  \alpha_{89}\,h_{1} h_{10} s^0_{30} s^0_{9}
+\alpha_{92}\,h_{2} h_{9} s^0_{30} s^0_{9} +\alpha_{94}\,s^0_{17}
s^0_{25} s^0_{30} s^0_{9} +\alpha_{95}\,s^0_{1} s^0_{18} s^0_{26}
s^0_{30} s^0_{9} +\alpha_{98}\,s^0_{18} s^0_{27} s^0_{30} s^0_{9}
\nonumber\\
\scriptstyle & \scriptstyle  & \scriptstyle  {} +\alpha_{103}\,s^0_{15} (s^0_{30})^2 s^0_{9}
+\alpha_{104}\,s^0_{16} (s^0_{30})^2 s^0_{9} +\alpha_{105}\,s^0_{20}
s^0_{30} s^0_{31} s^0_{9} +\alpha_{106}\,s^0_{21} s^0_{30} s^0_{31}
s^0_{9} \;,\\
\scriptstyle F_{19} & \scriptstyle  = & \scriptstyle  \alpha_{35}\,h_{1} h_{2} s^0_{30} s^0_{5} s^0_{6}
+\alpha_{36}\,s^0_{17} s^0_{18} s^0_{30} s^0_{5} s^0_{6}
+\alpha_{37}\,s^0_{20} s^0_{22} s^0_{30} s^0_{5} s^0_{6}
+\alpha_{38}\,s^0_{21} s^0_{22} s^0_{30} s^0_{5} s^0_{6}
+\alpha_{39}\,s^0_{20} s^0_{23} s^0_{30} s^0_{5} s^0_{6}
\nonumber\\
\scriptstyle & \scriptstyle  & \scriptstyle  {}
+\alpha_{40}\,s^0_{21} s^0_{23} s^0_{30} s^0_{5} s^0_{6}\;.
\end{eqnarray}
\end{subequations}
Here $\alpha_i$ denote superpotential coefficients. The $F$-term
equations, $F_i=0$, have trivial and non-trivial solutions. For
instance, \eqref{eq:F3} has the trivial solutions
\[ s^0_{26} ~=~0\quad\text{or}\quad s^0_{4} ~=~0\;,\]
as well as the non-trivial solution relating various VEVs,
\begin{equation}
s^0_{16}~=~\frac{-\alpha_{27} - \alpha_{28}\,s^0_{15}\,s^0_{26}
}{\alpha_{29}\,s^0_{26}}\;.
\end{equation}
The strategy is now to take the non-trivial solution and insert it
into the other equations. By doing so, one can trade the $F$-term
constraints for relations between the VEVs. We find that this
strategy is successful and we can satisfy all equations with
non-trivial $\widetilde{s}$ VEVs.

\subsubsection{Model 1, vacuum configuration B}

Now consider the `vacuum' configuration where the fields

\parbox{\linewidth}{
\begin{eqnarray}
\{\widetilde{s}_i\} = & \{\chi_{1}, \chi_{2}, \chi_{3}, \chi_{4},
h_{1}, h_{2}, h_{3}, h_{4}, h_{5}, h_{6}, h_{7}, h_{8}, h_{9},
h_{10}, s^0_{1},s^0_{2},s^0_{3},
s^0_{4}, &\nonumber\\
& \quad s^0_{5}, s^0_{6},s^0_{7},s^0_{8}, s^0_{9},s^0_{10},
s^0_{11},s^0_{12}, s^0_{13},s^0_{14},
s^0_{15},s^0_{16}, s^0_{17}, s^0_{18}, s^0_{20}, & \nonumber\\
 & \quad s^0_{21}, s^0_{22}, s^0_{23},s^0_{24}, s^0_{25}, s^0_{26}, s^0_{27},s^0_{28},s^0_{29},
s^0_{30}, s^0_{31},s^0_{32} \} &
\end{eqnarray}}
develop a VEV while the expectation values of all other fields
vanish.  Hence, in this vacuum configuration only one SM singlet VEV
is set to zero, i.e.  $\langle s^0_{19} \rangle = 0$. The emerging
effective theory has most properties identical to those in Model 1A:
\begin{enumerate}
\item the unbroken gauge symmetries are
\begin{equation}
 G_\mathrm{SM}\times G_\mathrm{hid}\;,
\end{equation}
where $G_\mathrm{hid}=\SU4$.
\item since \BmL\ is broken by two units, there is an effective
matter parity $\mathbb{Z}_2^{\cal M}$.
\item there is only one pair of Higgs candidates, $\phi_1$ and $\bar\phi_1$;
the $\mu$-term
\begin{equation}
 \mu~=~\left.\frac{\partial^2 W}{\partial\phi_1\,\partial\bar \phi_1}
 \right|_{\phi_1=\overline{\phi}_1=0}
\end{equation}
vanishes up to order $\widetilde{s}^6$, at which we work. That is,
there is one pair of massless Higgs doublets.
\item we check that the solution satisfies $F=0$ for all fields and in addition,
switching on $\{\widetilde{s}_i\}$-fields allows us to cancel the FI
term without inducing $D$-terms.
\item all vector-like exotics decouple (cf.\ Appendix~\ref{sec:exoticsMasses1b}).
\item neutrino masses are suppressed via the see-saw mechanism.
\end{enumerate}
However, the detailed form of the Yukawa and exotic mass matrices
has changed.

\subsubsection*{Charged fermion Yukawa matrices}

The charged fermion Yukawa matrices are
\begin{eqnarray}
Y_u&=& \left(
\begin{array}{ccc}
 \widetilde{s}^5 & \widetilde{s}^5 & \widetilde{s}^5 \\
 \widetilde{s}^5 & \widetilde{s}^5 & \widetilde{s}^5 \\
 \widetilde{s}^6 & \widetilde{s}^6 & 1
\end{array}
\right)\;,\quad Y_d~=~ \left(
\begin{array}{ccc}
 \widetilde{s}^5 & \widetilde{s}^5 & 0  \\
 \widetilde{s}^5 & \widetilde{s}^5 & 0  \\
 \widetilde{s}^6 & \widetilde{s}^6 & 0
\end{array}
\right)\;,\quad Y_e~=~ \left(
\begin{array}{ccc}
 \widetilde{s}^5 & \widetilde{s}^5 & \widetilde{s}^6 \\
 \widetilde{s}^5 & \widetilde{s}^5 & \widetilde{s}^6 \\
 \widetilde{s}^6 & \widetilde{s}^6 & 0
\end{array}
\right)\;.\nonumber\\
& &
\end{eqnarray}
The up-type quark Yukawa matrix is given directly in terms of the
coupling of the up-type Higgs to the three $q$ and $\bar u$ fields.
The down-type quark and charged lepton Yukawa matrices are obtained
by integrating out a pair of vector-like $d$- and $\bar d$-quarks
and $\ell$- and $\bar\ell$-fields, respectively. We find that (just
as in Model 1A) the up and charged lepton Yukawa matrices have rank
3, while the down quark Yukawa matrix has only rank 2, at this order
in $\tilde s$ singlets. In fact, to this order in SM singlet fields,
the superpotential does not couple two right-handed down quarks,
$\bar d_{3,4}$, to the quark doublets. This is because $\bar
d_{3,4}$ are in the $T_4$ twisted sector.  However, we have verified
that some of the zeros in $Y_d$ get filled in at higher orders and
at order 8 $Y_d$ has rank 3. Note that in this vacuum configuration
the Yukawa matrices retain a form consistent with the underlying
$D_4$ family symmetry.\footnote{The $D_4$ family symmetry is a
consequence of the space group selection rules and the geometry of
the $\SO4$ torus (Fig. \ref{fig:T1})
\cite{Kobayashi:2004ya,Kobayashi:2006wq}. States sitting at the two
vertical fixed points on the $\SO4$ torus transform as doublets
under $D_4$.}

\subsubsection*{Neutrino matrices}\label{sec:NeutrinoMasses1b}

We have checked that all right-handed neutrinos obtain mass in this
vacuum configuration.   Thus the See-Saw mechanism works exactly as
in Model 1A, although the detailed forms of the matrices differ. For
more details, see Webpage \cite{WebTables:2006ml}.

\subsubsection*{Dimension 5 baryon and lepton number violating operators}

We further analyzed the question of dimension 5 proton decay
operators. We find that both $q\,q\,q\,\ell$ and $\bar u\,\bar
u\,\bar d\,\bar e$ appear at order $\widetilde{s}^6$. They are also
generated by integrating out the heavy exotics.  For example, the
following couplings exist
\begin{equation}
  q_1 \,\ell_1\,\bar\delta_4\,,\; q_1 \,\ell_1\,\bar\delta_5\,,\; q_2
\,\ell_2\,\bar\delta_4\,,\;q_2
  \,\ell_2\,\bar\delta_5\,,\;
  q_1 \,q_1\,\delta_4\,,\;  q_1 \,q_1\,\delta_5\,,\;q_2
\,q_2\,\delta_4\,,\;q_2 \,q_2\,\delta_5\;.
\label{eq:trilinear_D5_operators}
\end{equation}
Hence integrating out the states  $\bar\delta_i, \ \delta_i$
produces dangerous dimension 5 operators. These must be sufficiently
suppressed to be consistent with present bounds on proton decay
\cite{Hinchliffe:1992ad,Dermisek:2000hr}. We have verified that, for
some particular $\widetilde s$ VEVs, it is possible to suppress the
$q\,q\,q\,\ell$ operators induced by the trilinear
couplings~\eqref{eq:trilinear_D5_operators}. However, higher order
couplings also introduce baryon and lepton number violating
operators. We have not been able to identify a suppression mechanism
for such operators yet.

\subsubsection*{$\boldsymbol{\mu}$-term and Minkowski space}

As in Model 1A, requiring a Minkowski vacuum puts a constraint on
the total superpotential which includes contributions from charged
matter and moduli.  An interesting feature of the present vacuum
configuration is that the SM matter singlet superpotential to order
$\tilde s^6$ is of the form,
\begin{equation}
  W(\widetilde{s} ) = \sum_i  P_i(\widetilde{D}) \ \tilde
  P_i(\widetilde{s}) \ ,
\end{equation}
where $P_i$ are polynomials in SM singlet fields (the index $i$
labels a particular polynomial) which are either $D_4$ doublets,
which we now re-label as $\widetilde{D}$, or SM and $D_4$ singlets,
$\widetilde{s}$. In particular, the $D_4$ doublets which enter
$W(\widetilde{s})$ are
\begin{eqnarray}
\widetilde{D}_1 = (s^0_3, s^0_9) & \qquad & \widetilde{D}_2 = (s^0_4, s^0_{10}) \\
\widetilde{D}_3 = (s^0_5, s^0_{11}) & \qquad & \widetilde{D}_4 = (s^0_6, s^0_{12}) \nonumber \\
\widetilde{D}_5 = (s^0_7, s^0_{13}) & \qquad & \widetilde{D}_6 =
(s^0_8, s^0_{14}) \ . \nonumber
\end{eqnarray}
The polynomial in $D_4$ doublets is, to this order, quadratic in
doublets and is given by the trivial $D_4$ singlet scalar product,
for example,
\begin{equation} \widetilde{D}_1 \cdot \widetilde{D}_2 = (s^0_3 \ s^0_4 + s^0_9 \
s^0_{10}) \ .
\end{equation}
We then find (up to calculable dimensionful coefficients in units of
the string scale)
\begin{eqnarray}
W& =&\left(\widetilde{D}_1 \cdot \widetilde{D}_2\right)\ \left(
s^0_{26} + s^0_{29} + (s^0_{26} s^0_{26} + s^0_{26} s^0_{29} +
s^0_{29} s^0_{29}) (s^0_{15} + s^0_{16})\right) \nonumber \\ &&
{}+\left(\widetilde{D}_1 \cdot \widetilde{D}_6 + \widetilde{D}_2
\cdot \widetilde{D}_5\right)\ s^0_{30}\ \biggl[ s^0_{30} (s^0_{15} +
s^0_{16}) + s^0_{17} (s^0_{25} + s^0_{28}) \nonumber \\ && {}+
s^0_{18} (s^0_{24} + s^0_{27})+ s^0_{31} (s^0_{20} + s^0_{21}) +
s^0_{32} (s^0_{22} + s^0_{23})  \nonumber \\ && {}+ (s^0_{19} +
s^0_1 s^0_{18}+ s^0_2 s^0_{17}) (s^0_{26} + s^0_{29})+ h_1 (h_8 +
h_{10}) + h_2 (h_7 + h_9)\biggr] \nonumber \\ && {}+
\left(\widetilde{D}_3 \cdot \widetilde{D}_4\right)\ s^0_{19}
s^0_{30}\ \left( s^0_{17} s^0_{18} + h_1 h_2 + (s^0_{20} + s^0_{21})
(s^0_{22} + s^0_{23})\right) \; . \label{eq:W}
\end{eqnarray}
Thus, to order 6 in SM and hidden $\SU4$ singlets,  the polynomials
$P_i(\widetilde{D})$ are completely determined by the $D_4$
symmetry, while the polynomials $\tilde P_i(\widetilde{s})$ are
non-trivial for all $i$. One particular $F= D =0$ solution is given
by the roots of $\langle P_i(\widetilde{D}) \rangle = \langle \tilde
P_i(\widetilde{s}) \rangle = 0$ for all polynomials $i$.  Hence,
once again,
\begin{equation}
 \left\langle  W(\widetilde{s} ) \right\rangle~=~0 \;.
\end{equation}
Therefore, the total superpotential is given solely by its
non--perturbative part. This is expected to be very small and thus a
small gravitino mass and a small cosmological constant can in
principle be achieved.

In addition, just as in Model 1A, the $\mu$-term contains all terms
present in $W(\widetilde{s} )$ and to order 6 in SM singlets we have
$\mu = 0$, when $W=0$. This means that the $\mu$-term is of the
order of the expectation value of $W$, i.e.\ the gravitino mass.

\subsection{Model 2}

The model is defined by the shift and Wilson lines given in Appendix
F.  It was already included as an example in ML
\cite{Lebedev:2006kn} and \cite{Buchmuller:2007zd}.

The gauge group after compactification is
\begin{equation}
G ~=~[\SU3\times\SU2]\times[\SO8\times\SU2]\times\U1^8\;.
\end{equation}

As before, we are able to define a suitable $B-L$ generator,
\begin{equation}\label{eq:tBmL2}
 \mathsf{t}_{B-L}~=~
 \left(
 1 , 1 , 0 , 0 , 0 , -\frac{2}{3} , -\frac{2}{3} , -\frac{2}{3} \right)\,
 \left(
 \frac{1}{2} , \frac{1}{2} , 0 , \frac{1}{2} , \frac{1}{2} , 0 , 0 , 0
 \right)\;
\end{equation}
with two important properties (cf.\ Table~\ref{Tab:spectrum}):
\begin{itemize}
\item  the spectrum includes 3 generations of quarks and leptons plus vector-like exotics
with respect to $G_\mathrm{SM}\times\U1_{\BmL}\;,$ and
\item there are SM singlets with \BmL\ charge $\pm2$.
\end{itemize}

\begin{table}[h]
\begin{center}
\begin{tabular}{|c|l|l|c|c|l|l|}
\hline
\# & irrep & label & & \# & irrep & label\\
\hline
 3 &
$\left(\boldsymbol{3},\boldsymbol{2};\boldsymbol{1},\boldsymbol{1}\right)_{(1/6,1/3)}$
 & $q_i$
 & &
 3 &
$\left(\overline{\boldsymbol{3}},\boldsymbol{1};\boldsymbol{1},\boldsymbol{1}\right)_{(-2/3,-1/3)}$
 & $\bar u_i$
 \\
 3 &
$\left(\boldsymbol{1},\boldsymbol{1};\boldsymbol{1},\boldsymbol{1}\right)_{(1,1)}$
 & $\bar e_i$
 & &
 4 &
$\left(\boldsymbol{1},\boldsymbol{2};\boldsymbol{1},\boldsymbol{1}\right)_{(0,*)}$
 & $m_i$
 \\
 4 &
$\left(\overline{\boldsymbol{3}},\boldsymbol{1};\boldsymbol{1},\boldsymbol{1}\right)_{(1/3,-1/3)}$
 & $\bar d_i$
 & &
 1 &
$\left(\boldsymbol{3},\boldsymbol{1};\boldsymbol{1},\boldsymbol{1}\right)_{(-1/3,1/3)}$
 & $d_i$
 \\
 4 &
$\left(\boldsymbol{1},\boldsymbol{2};\boldsymbol{1},\boldsymbol{1}\right)_{(-1/2,-1)}$
 & $\ell_i$
 & &
 1 &
$\left(\boldsymbol{1},\boldsymbol{2};\boldsymbol{1},\boldsymbol{1}\right)_{(1/2,1)}$
 & $\bar \ell_i$
 \\
 4 &
$\left(\boldsymbol{1},\boldsymbol{2};\boldsymbol{1},\boldsymbol{1}\right)_{(-1/2,0)}$
 & $\phi_i$
 & &
 4 &
$\left(\boldsymbol{1},\boldsymbol{2};\boldsymbol{1},\boldsymbol{1}\right)_{(1/2,0)}$
 & $\bar \phi_i$
 \\
 3 &
$\left(\overline{\boldsymbol{3}},\boldsymbol{1};\boldsymbol{1},\boldsymbol{1}\right)_{(1/3,2/3)}$
 & $\bar\delta_i$
 & &
 3 &
$\left(\boldsymbol{3},\boldsymbol{1};\boldsymbol{1},\boldsymbol{1}\right)_{(-1/3,-2/3)}$
 & $\delta_i$
 \\
20 &
$\left(\boldsymbol{1},\boldsymbol{1};\boldsymbol{1},\boldsymbol{1}\right)_{(1/2,*)}$
 & $s^+_i$
 & &
 20 &
$\left(\boldsymbol{1},\boldsymbol{1};\boldsymbol{1},\boldsymbol{1}\right)_{(-1/2,*)}$
 & $s^-_i$
 \\
 15 &
$\left(\boldsymbol{1},\boldsymbol{1};\boldsymbol{1},\boldsymbol{1}\right)_{(0,1)}$
 & $\bar n_i$
 & &
 12 &
$\left(\boldsymbol{1},\boldsymbol{1};\boldsymbol{1},\boldsymbol{1}\right)_{(0,-1)}$
 & $n_i$
 \\
 3 &
$\left(\boldsymbol{1},\boldsymbol{1};\boldsymbol{1},\boldsymbol{2}\right)_{(0,1)}$
 & $\bar \eta_i$
 & &
 3 &
$\left(\boldsymbol{1},\boldsymbol{1};\boldsymbol{1},\boldsymbol{2}\right)_{(0,-1)}$
 & $\eta_i$
 \\
 20 &
$\left(\boldsymbol{1},\boldsymbol{1};\boldsymbol{1},\boldsymbol{2}\right)_{(0,0)}$
 & $h_i$
 & &
 2 &
$\left(\boldsymbol{1},\boldsymbol{2};\boldsymbol{1},\boldsymbol{2}\right)_{(0,0)}$
 & $y_i$
 \\
 2 &
$\left(\boldsymbol{1},\boldsymbol{1};\boldsymbol{1},\boldsymbol{2}\right)_{(1/2,1)}$
 & $x^+_i$
 & &
 2 &
$\left(\boldsymbol{1},\boldsymbol{1};\boldsymbol{1},\boldsymbol{2}\right)_{(-1/2,-1)}$
 & $x^-_i$
 \\
 2 &
$\left(\boldsymbol{1},\boldsymbol{1};\boldsymbol{1},\boldsymbol{1}\right)_{(0,\pm2)}$
 & $\chi_i$
 & &
 18 &
$\left(\boldsymbol{1},\boldsymbol{1};\boldsymbol{1},\boldsymbol{1}\right)_{(0,0)}$
 & $s^0_i$
 \\
 4 &
$\left(\overline{\boldsymbol{3}},\boldsymbol{1};\boldsymbol{1},\boldsymbol{1}\right)_{(-1/6,*)}$
 & $\bar v_i$
 & &
 4 &
$\left(\boldsymbol{3},\boldsymbol{1};\boldsymbol{1},\boldsymbol{1}\right)_{(1/6,*)}$
 & $v_i$
 \\
 2 &
$\left(\boldsymbol{1},\boldsymbol{1};\boldsymbol{8},\boldsymbol{1}\right)_{(0,-1/2)}$
 & $f_i$
 & &
 2 &
$\left(\boldsymbol{1},\boldsymbol{1};\boldsymbol{8},\boldsymbol{1}\right)_{(0,1/2)}$
 & $\bar f_i$
 \\
 5 &
$\left(\boldsymbol{1},\boldsymbol{1};\boldsymbol{8},\boldsymbol{1}\right)_{(0,0)}$
 & $w_i$
 & &
  &
 &
 \\
\hline
\end{tabular}
\end{center}
\caption{Spectrum. The quantum numbers under
$\SU3\times\SU2\times[\SO8\times\SU2']$ are shown in boldface; hypercharge and \BmL\ charge
appear as subscript.  Note that the states $s_i^\pm$, $m_i$ and $v_i$ have
different B-L charges for different $i$, which we do not explicitly list.}
\label{Tab:spectrum}
\end{table}

Consider a `vacuum' configuration where the fields
\begin{eqnarray}
 \{\widetilde{s}_i\}
 ~=&
 \{
 \chi_{1}, \chi_{2}, s^0_{3}, s^0_{5}, s^0_8, s^0_9, s^0_{12}, s^0_{15}, s^0_{16}, s^0_{22}, s^0_{24}, s^0_{35},
 s^0_{41}, s^0_{43}, s^0_{46}, & \nonumber \\ & h_{2}, h_3, h_{5}, h_{9}, h_{13}, h_{14}, h_{20},
 h_{21}, h_{22}
 \} & \label{eq:stilde}
\end{eqnarray}
develop a VEV while the expectation values of all other fields
vanish. The emerging effective theory has the following properties:
\begin{enumerate}
\item the unbroken gauge symmetries are
\begin{equation}
 G_\mathrm{SM}\times G_\mathrm{hid}\;,
\end{equation}
where $G_\mathrm{hid}=\SO8$.
\item since \BmL\ is broken by two units, there is an effective matter parity $\mathbbm{Z}_2^{\cal M}$.
\item the Higgs mass terms are
\begin{equation}
 \bar\phi_i\,(\mathcal{M}_{\bar\phi\phi})_{ij}\,\phi_j\;,
 \quad\text{where}\quad
 \mathcal{M}_{\bar\phi\phi}~=~
 \left(
 \begin{array}{cccc}
 \widetilde{s}^4 & 0 & 0 & \widetilde{s} \\
 \widetilde{s} & \widetilde{s}^3 & \widetilde{s}^3 & \widetilde{s}^6 \\
 \widetilde{s}^5 & 0 & 0 & \widetilde{s}^3 \\
 \widetilde{s} & 0 & 0 & \widetilde{s}^3
 \end{array}
 \right)\;.
\end{equation}
The up-type Higgs $h_u$ is a linear combination of $\bar\phi_1$,
$\bar\phi_3$ and $\bar\phi_4$,
\begin{equation}
 h_u~\sim~\widetilde{s}^{2}\bar\phi_1+\bar\phi_3+\widetilde{s}^4\,\bar\phi_4\;,
\end{equation}
while the down-type Higgs is composed out of $\phi_2$ and $\phi_3$,
\begin{equation}
 h_d~\sim~\,\phi_2+\phi_3\;.
\end{equation}
The vacuum configuration is chosen such that the $\mu$-term, being defined as
the smallest eigenvalue of $\mathcal{M}_{\bar\phi\phi}$,
\begin{equation}
 \mu~=~\left.\frac{\partial^2 W}{\partial h_d\,\partial h_u}\right|_{h_u=h_d=0}
\end{equation}
vanishes up to order $\widetilde{s}^6$, at which we work.
\item we check that
switching on $\{\widetilde{s}_i\}$-fields allows us to cancel the FI
term without inducing $D$-terms (cf.\ Appendix~\ref{sec:Dflat}).
\item all exotics decouple (cf.\ Appendix~\ref{sec:exoticsMasses2}).
\item neutrino masses are suppressed via the see-saw mechanism.
\end{enumerate}

Thus, again we have obtained a supersymmetric vacuum with the
precise matter content of the MSSM and R parity.

\subsubsection*{Charged fermion Yukawa matrices}

The up-Higgs Yukawa couplings decompose into
\begin{equation}
 W_\mathrm{Yukawa}~\supset~\sum\limits_{k=1}^4
 (Y_u)_{ij}^{(k)}\,q_i\,\bar u_j\,\bar\phi_k\,,
\end{equation}
where
\begin{subequations}
\begin{eqnarray}
 Y_u^{(1)}~=~
 \left(
 \begin{array}{ccc}
 0 & 0 & \widetilde{s}^6 \\
 0 & 0 & \widetilde{s}^6 \\
 \widetilde{s}^3 & \widetilde{s}^3 & 1
 \end{array}
 \right),\; &
 Y_u^{(2)}~=~
 \left(
 \begin{array}{ccc}
 0 & 0 & 0 \\
 0 & 0 & 0 \\
 0 & 0 &  \widetilde{s}^6
 \end{array}
 \right),\; & \\
 Y_u^{(3)}~=~
 \left(
 \begin{array}{ccc}
 0 & 0 & 0 \\
 0 & 0 & 0 \\
 0 & 0 & \widetilde{s}^6
 \end{array}
 \right),\; &
 Y_u^{(4)}~=~
 \left(
 \begin{array}{ccc}
 0 & 0 & 0 \\
 0 & 0 & 0 \\
 0 & 0 & \widetilde{s}^6
 \end{array}
 \right)\; & .  \nonumber
\end{eqnarray}
\end{subequations}

Thus, the physical $3\times3$ up-Higgs Yukawa matrix is
\begin{equation}
 Y_u~\sim~\widetilde{s}^2\,Y_u^{(1)}+Y_u^{(3)}+\widetilde{s}^4\,Y_u^{(4)}~=~
 \left(
 \begin{array}{ccc}
 0 & 0 & \widetilde{s}^8 \\
 0 & 0 & \widetilde{s}^8 \\
 \widetilde{s}^5 & \widetilde{s}^5 & \widetilde{s}^2
 \end{array}
 \right)\,.
\end{equation}
Note that due to the Higgs mixing the top quark Yukawa coupling for
this vacuum configuration is given by $\widetilde{s}^2$. Thus the
corresponding $\widetilde{s}$ VEVs are required to be quite large.

The down-Higgs Yukawa couplings decompose into
\begin{equation}
 W_\mathrm{Yukawa}~\supset~\sum\limits_{k=1}^4
 (Y_d)_{ij}^{(k)}\,q_i\,\bar d_j\,\phi_k\,,
\end{equation}
where
\begin{subequations}
\begin{eqnarray}
 Y_d^{(1)}~=~
 \left(
 \begin{array}{cccc}
 \widetilde{s}^4 & \widetilde{s}^4 & \widetilde{s}^5 & \widetilde{s}^5 \\
 \widetilde{s}^4 & \widetilde{s}^4 & \widetilde{s}^5 & \widetilde{s}^5 \\
 \widetilde{s}^5 & \widetilde{s}^5 & \widetilde{s}^6 & \widetilde{s}^6
 \end{array}
 \right),\; &
 Y_d^{(2)}~=~
 \left(
 \begin{array}{cccc}
 1 & \widetilde{s}^4 & 0 & 0 \\
 \widetilde{s}^4 & 1 & 0 & 0 \\
 \widetilde{s} & \widetilde{s} & 0 & 0
 \end{array}
 \right),\; & \\
 Y_d^{(3)}~=~
 \left(
 \begin{array}{cccc}
 1 & \widetilde{s}^4 & 0 & 0 \\
 \widetilde{s}^4 & 1 & 0 & 0 \\
 \widetilde{s} & \widetilde{s} & 0 & 0
 \end{array}
 \right),\; &
 Y_d^{(4)}~=~0 \;& .  \nonumber
\end{eqnarray}
\end{subequations}

The physical $3\times3$ down-Higgs Yukawa matrix emerges by
integrating out a pair of vector-like $d-$ and $\bar d-$quarks,
\begin{equation}
 Y_d~=~
 \left(
 \begin{array}{ccc}
 1 & \widetilde{s}^3 & 0 \\
 1 & \widetilde{s}^3 & 0 \\
 \widetilde{s} & \widetilde{s}^4 & 0
 \end{array}
 \right)\,.
\end{equation}
We note that both the up and down quarks are massless at order 6 in
SM singlets. However, we have checked that the up quark becomes
massive at order 7 and the down quark gets a mass at order 8.

The charged lepton Yukawa couplings decompose into
\begin{equation}
 W_\mathrm{Yukawa}~\supset~\sum\limits_{k=1}^4
 (Y_e)_{ij}^{(k)}\,\ell_i\,\bar e_j\,\phi_k\,,
\end{equation}
where
\begin{subequations}
\begin{eqnarray}
 Y_e^{(1)}~=~
 \left(
 \begin{array}{ccc}
 \widetilde{s}^4 & \widetilde{s}^4 & \widetilde{s}^5 \\
 \widetilde{s}^4 & \widetilde{s}^4 & \widetilde{s}^5 \\
 0 & 0 & 0 \\
 0 & 0 & 0
 \end{array}
 \right),\; &
 Y_e^{(2)}~=~
 \left(
 \begin{array}{ccc}
 1 & \widetilde{s}^4 & \widetilde{s}\\
 \widetilde{s}^4 & 1 & \widetilde{s}\\
 0 & 0 & \widetilde{s}^6  \\
 0 & 0 & \widetilde{s}^6
 \end{array}
 \right),\; & \\
 Y_e^{(3)}~=~
 \left(
 \begin{array}{ccc}
 1 & \widetilde{s}^4 & \widetilde{s}\\
 \widetilde{s}^4 & 1 & \widetilde{s}\\
 0 & 0 & \widetilde{s}^6  \\
 0 & 0 & \widetilde{s}^6
 \end{array}
 \right),\; &
 Y_e^{(4)}~=~
 \left(
 \begin{array}{ccc}
 0 &  0 &  \widetilde{s}^5 \\
 0 &  0 &  \widetilde{s}^5 \\
 0 &  0 &  \widetilde{s}^6 \\
 0 &  0 &  \widetilde{s}^6
 \end{array}
 \right)\; & . \nonumber
\end{eqnarray}
\end{subequations}
The physical $3\times3$ matrix emerges by integrating out a pair of
vector-like $\ell-$ and $\bar\ell-$leptons,

\begin{equation}
 Y_e~=~
 \left(
 \begin{array}{ccc}
 1 & 1 & \widetilde{s} \\
 \widetilde{s} & \widetilde{s} & \widetilde{s}^2 \\
 0 &  0 &  \widetilde{s}^6 \\
 \end{array}
 \right)\,.
\end{equation}

\subsubsection*{Neutrino masses} \label{sec:NeutrinoMasses}

We consider vacua where $\SU2'$ is broken. This means that the
$\eta_i$ and $\bar\eta_i$ give rise to further SM singlets with
$q_{\BmL}=\pm1$,
\begin{equation}
 \bar\eta_1
 ~=~
 \left(\begin{array}{c}
 \bar n_{16}\\
 \bar n_{17}\end{array}\right)\;,\dots
 \bar\eta_3
 ~=~
 \left(\begin{array}{c}
 \bar n_{20}\\
 \bar n_{21}\end{array}\right)
 \quad\text{and}\quad
 \eta_1
 ~=~
 \left(\begin{array}{c}
  n_{13}\\
  n_{14}\end{array}\right)\;,\dots
 \eta_3
 ~=~
 \left(\begin{array}{c}
  n_{17}\\
  n_{18}\end{array}\right)
 \;.
\end{equation}

The dimensions of the ``right-handed" neutrino mass matrices are
\begin{eqnarray}
\mathcal{M}_{nn} & = & 18 \times 18\;,
\\
\mathcal{M}_{n\bar n} & = & 18 \times 21\;,
\\
\mathcal{M}_{\bar n\bar n} & = & 21 \times 21\;,
\end{eqnarray}
with the neutrino mass matrix given by
\begin{equation}
 \mathcal{M}_{\bar\nu\bar\nu}~=~\left(\begin{array}{cc}
 \mathcal{M}_{\bar n\bar n} & \mathcal{M}_{n\bar n}^T\\
 \mathcal{M}_{n\bar n} & \mathcal{M}_{nn}
 \end{array}\right)\;.
\end{equation}
We have checked that it has full rank.

The neutrino Yukawa couplings decompose into
\begin{equation}
 W_\mathrm{Yukawa}~\supset~\sum\limits_{k=1}^4
 (Y_n)_{ij}^{(k)}\,\ell_i\,n_j\,\bar\phi_k
 +(Y_{\bar n})_{ij}^{(k)}\,\ell_i\,\bar n_j\,\bar\phi_k ,
\end{equation}
where $ Y_n^{(1)}, Y_n^{(2)}$ are non-vanishing $4 \times 18$
matrices, $Y_n^{(k>2)} =  0$ and  $Y_{\bar n}^{(1)}, Y_{\bar
n}^{(2)},  Y_{\bar n}^{(3)}, Y_{\bar n}^{(4)}$ are non-vanishing $4
\times 21$ matrices.  The effective neutrino mass matrix obtained as
\begin{equation}
 \kappa~=~Y_\nu\,\mathcal{M}_{\bar\nu\bar\nu}^{-1}\,Y_\nu^{T}\;,
\end{equation}
where $Y_\nu~=~(Y_n,Y_{\bar n})$ and $\kappa$ has non-zero
determinant. See Webpage \cite{WebTables:2006ml} for details.

\subsubsection*{Dimension 5 baryon and lepton number violating operators}

We have looked for effective dimension 5 baryon and lepton number
violating operators in this model.   We find that to order $\tilde
s^6$  no such operators exist.   However, these operators can be
generated once the exotics  $\delta_i, \ \bar \delta_i$ are
integrated out. Fortunately, a clever choice of VEVs for the fields
$\{\widetilde s_i\}$ can guarantee sufficient suppression of all
induced $q\,q\,q\,\ell$ operators, consistent with current bounds on
proton decay~\cite{Hinchliffe:1992ad,Dermisek:2000hr}.

\subsubsection*{$\boldsymbol{\mu}$-term and Minkowski space}

Unlike in the previous model, there is no relation between the
$\mu$--term and $ W(\tilde s )$. This is because the Higgs doublets
do not come entirely from the untwisted sector. Requiring
spontaneous SUSY breaking in a Minkowski vacuum puts a constraint on
the moduli VEVs. Fine--tuning is likely to be necessary to obtain a
realistic  gravitino mass as well as a small cosmological constant.

\section{Conclusions and Discussion \label{sec:conclusion}}

In this paper we have described the construction of heterotic MSSMs
with R parity. Our setup is based on a particular \Z6-II orbifold
with an SO(10) local GUT structure.  In the first part of the paper
we have obtained 218 models  with the MSSM gauge group structure, 3
light families and vector-like exotics. We show that all the
vector-like exotics can decouple along $D$--flat directions for 190
of these models.\footnote{In this analysis, we have taken into
account superpotential terms up to order 6 in SM singlets. At higher
orders, we expect more models to be retained.} The total number of
inequivalent models with SO(10) shifts and 2 Wilson lines is
$3\cdot10^4$. Hence 0.6\,\% of our total model set are MSSM
candidates. This can be compared with D brane constructions where
the probability of getting MSSM-like models is much less than
$10^{-9}$ or Gepner orientifold constructions where this probability
is $10^{-14}$.

In the second part of the paper we go further down the road towards
the MSSM. We define a successful strategy for obtaining models with
an exact R parity. We find 87 models which have a renormalizable top
Yukawa coupling.  We identify 15 models with an exact R parity, no
light exotics or U(1) gauge bosons and an order one top quark Yukawa
coupling.\footnote{The number 15 is a lower bound, since our search
is based on a specific strategy related to $B-L$ symmetry.
Furthermore, more models are retained if we do not insist on having
a renormalizable top Yukawa coupling. Also one can drop the strict
constraint that exotics be vector-like with respect to $B-L$. For
example, two exotics $x, \ \bar x$ with $B-L$ charge -1 can get mass
from a SM singlet VEV with charge +2.}

We present two explicit ``benchmark" examples satisfying the following criteria:
\begin{itemize}
 \item  MSSM spectrum below the string scale -
  \subitem   all exotics decouple;
  \subitem   one pair of light Higgs doublets;
  \subitem   top quark Yukawa coupling of order 1;
  \subitem   non-trivial Yukawa matrices for charged fermions;
  \subitem   See-Saw mechanism  for neutrinos;
\item  an exact R parity.
\end{itemize}

The two examples have different phenomenological properties such as
different structures of the Yukawa coupling matrices and dimension 5
operators.  In particular,  the Yukawa matrices $Y_d$ and $Y_e$ have
more non-vanishing entries in Model 1B than in Model 1A.  In both
Models 1A/B the lightest down type quark is massless at order 6 in
SM singlets and becomes massive at order 8. The top Yukawa coupling
is order one in Models 1A/B, while it is order $\widetilde{s}^2$ in
Model 2. This is due to Higgs doublet mixing in the latter.  In
Model 2, both the up and down quarks are massless at order 6 in SM
singlets. However, the up quark becomes massive at order 7 and the
down quark gets a mass at order 8.

An interesting feature of Model 1 is that there is a correlation
between the $\mu$ term and the expectation value of the
superpotential. In fact the pair of Higgs fields are the only
vector-like fields whose mass is correlated with the expectation
value of the superpotential, while all exotics can consistently get
mass with $W=0$.  This provides a novel, stringy solution to the
MSSM $\mu$ problem.  Indeed, in Models 1A/B the vacuum expectation
value of the superpotential and $\mu$ both vanish at order 6 in SM
singlets. Thus, neglecting non-perturbative effects, this model
leads to a supersymmetric Minkowski vacuum with $\mu=0$. One expects
that when non-perturbative effects (hidden sector gaugino
condensation) are taken into account, supersymmetry is broken at a
hierarchically small scale and, because of the correlation between
$\mu$ and $\langle W\rangle$, $\mu$ is of order the gravitino mass.
In Model 2, on the other hand, the superpotential does not vanish in
this limit and inclusion of non-perturbative contributions to the
superpotential is necessary.

Dimension 5 baryon and lepton number violating operators come from
two sources.  They are generated in the superpotential to some order
in SM singlets.  They may also be generated when integrating out
heavy exotics.  In Models 1A/B the direct dimension 5 operators
appear at order $\widetilde{s}^6$, while in Model 2 they do not
appear at this order.   In addition, in Models 1A/B and Model 2
dimension 5 operators appear when integrating out heavy exotics. In
Model 2 these can be sufficiently suppressed with some fine-tuning.

There are some phenomenological issues that we have not addressed in
this paper. In particular, we have not studied precision gauge
coupling unification. Although hypercharge is normalized as in 4D
GUTs thus allowing gauge coupling unification in the first
approximation, there are various corrections that can be important.
First, a detailed analysis would require the calculation of string
threshold corrections in the presence of discrete Wilson lines.
However in specific cases these corrections are known to be small
\cite{Mayr:1993kn}. Second, there are corrections from the
vector--like exotic states. It is possible that precision gauge
coupling unification may require anisotropic compactifications,
leading to an effective orbifold GUT
\cite{Witten:1996mz,Ibanez:1992hc,Kobayashi:2004ya,Hebecker:2004ce}.

Another issue concerns proton stability. The examples we studied are challenged
by the presence of dimension 5 proton decay operators. Their suppression may
require additional (discrete) symmetries. There are also dimension 6 operators,
generated by GUT gauge boson exchange, which we have not discussed.

Finally, there are the usual questions of moduli stabilization and supersymmetry
breakdown in a Minkowski vacuum. Some of them we discussed previously in
\cite{Lebedev:2006tr}. We have not addressed all of these issues here.  On the
other hand, it is clear that if given the freedom of arbitrarily tuning moduli
VEVs we are not able to find the MSSM, the whole approach would be futile.
However, with a number of MSSM candidates in this fertile patch of the
landscape, it is now imperative to tackle the hard problems just mentioned.

\subsubsection*{Acknowledgments}

We would like to thank K.S.~Choi and J.~Gray and T.~Kobayashi for
discussions. O.L., S.R., S.R-S., P.V. and A.W. would like to thank
TUM for hospitality and support. S.R. and A.W. also thank Bonn
University for hospitality and support.  M.R. would like to thank
the Summer Institute 2007 (held at Fuji-Yoshida) and the Aspen
Center for Physics for hospitality and support. This research was
supported by the DFG cluster of excellence Origin and Structure of
the Universe, the European Union 6th framework program
MRTN-CT-2004-503069 "Quest for unification", MRTN-CT-2004-005104
"ForcesUniverse", MRTN-CT-2006-035863 "UniverseNet" and
SFB-Transregios 27 "Neutrinos and Beyond" and 33 "The Dark Universe"
by Deutsche Forschungsgemeinschaft (DFG). S.R. and A.W. received
partial support from DOE grant DOE/ER/01545-874.

\appendix

\section{Physical states and string selection rules}
In this appendix, we discuss how to build consistent physical
states. Furthermore, we list the string selection rules used in this
work. Finally, we comment on an additional selection rule present in
the literature: the $\gamma$ rule. We find that our construction of
physical states is useful in order to apply the $\gamma$ rule
correctly. It turns out that, in contrast to previous statements,
the $\gamma$ rule does not further constrain allowed couplings.

\subsection{Physical states}
\label{sec:physstates}

An element of the \emph{space group} $g = (\theta^k, n_\alpha
e_\alpha)\in S$, where $\theta $ is the twist and $e_\alpha$ are the
lattice basis vectors, corresponds to a boundary condition of a
closed string~\cite{Dixon:1986jc, Dixon:1985jw}. For $k = 0$ ($k
\neq 0$), the string is named \emph{untwisted string} (\emph{twisted
string}). Focusing on its bosonic degrees of freedom in the six
extra dimensions, the boundary condition reads
\begin{equation}
\label{eqn:boundary} X(\tau, \sigma + 2\pi)~=~ g\,X(\tau, \sigma)\;,
\end{equation}
where $g$ is called the \emph{constructing element} of the closed
string. For each constructing element $g$, there exists a
corresponding Hilbert space $\mathcal{H}_g$ of physical states.
Using a mode expansion for $X(\tau, \sigma)$, the general solutions
of the string equation of motion with boundary condition
Eq.~\eqref{eqn:boundary} can be written down. From these solutions
one finds that twisted strings are localized at the \emph{
fixed--point} $f_g \in \mathbbm{R}^6$ corresponding to $g$ (i.e.\
$\theta^k f_g +  n_\alpha e_\alpha = f_g$). Furthermore, their
quantization leads to the mass equation for left--movers (and the
mass equation for right--movers is derived analogously). Focusing on
the massless case, the solutions are denoted by\footnote{In this
discussion we disregard oscillator states. Their inclusion is
straightforward and does not change our conclusions.}
\begin{equation}
|q_\mathrm{sh}\rangle_\mathrm{R} \otimes
|P_\mathrm{sh}\rangle_\mathrm{L}\;,
\end{equation}
with \emph{shifted momenta} $q_\mathrm{sh} \equiv q + v_g$ and
$P_\mathrm{sh} \equiv P + V_g$, where $q$ and $P$ lie in the SO(8)
weight lattice and $E_8 \times E_8$ root lattice, respectively. The
local twist and shift corresponding to the space group element $g =
(\theta^k, n_\alpha e_\alpha)$ are defined by $v_g \equiv kv$ and
$V_g \equiv k V + n_\alpha W_\alpha$, respectively. Since the string
is completely specified by its constructing element $g$ and its
left- and right--moving shifted momenta $P_\mathrm{sh}$ and
$q_\mathrm{sh}$, we write down a first ansatz for a physical state:
\begin{equation}
\label{eqn:physicalstate1} |\text{phys}\rangle~\sim~
|q_\mathrm{sh}\rangle_\mathrm{R} \otimes
|P_\mathrm{sh}\rangle_\mathrm{L} \otimes |g\rangle
\end{equation}

Up to now it is not guaranteed that a physical state is actually
compatible with the orbifold. To ensure this compatibility,
invariance of $|\text{phys}\rangle$ under the action of all elements
of the orbifold group $O \subset S \otimes G$ must be imposed ($G$
is the embedding of $S$ into the gauge degrees of freedom and is
called the \emph{gauge twisting group}). To do so,
Eq.~\eqref{eqn:boundary} is multiplied by an arbitrary element $h =
(\theta^l, m_\alpha e_\alpha) \in S$:
\begin{eqnarray}
                 h\,X(\tau, \sigma + 2\pi) & = & h\,g\,X(\tau, \sigma) \\
\Leftrightarrow  h\,X(\tau, \sigma + 2\pi) & = &
h\,g\,h^{-1}\,h\,X(\tau, \sigma) \label{eqn:boundarytimesh}
\end{eqnarray}
Furthermore, the transformation properties of left- and
right--movers under $h$ are:
\begin{equation}
\label{eqn:trafophase} |q_\mathrm{sh}\rangle_\mathrm{R} \otimes
|P_\mathrm{sh}\rangle_\mathrm{L} \xrightarrow{h} \Phi \;
|q_\mathrm{sh}\rangle_\mathrm{R} \otimes
|P_\mathrm{sh}\rangle_\mathrm{L}\;,
\end{equation}
where\footnote{Here, we set $\Phi_\text{vac} = 1$ as discussed
in~\cite{Ploger:2007iq}.}
\begin{equation}
\Phi \equiv e^{2\pi \,\I\, \left[q_\mathrm{sh}\cdot v_h -
P_\mathrm{sh}\cdot V_h\right]}\;.
\end{equation}
Now, we can distinguish two cases:

\subsubsection*{Commuting elements: $\boldsymbol{[h,g] = 0}$}

First, let us consider the transformation property of
$|\text{phys}\rangle$ with respect to a commuting element $h$. In
this case, Eq.~\eqref{eqn:boundarytimesh} yields
\begin{equation}
h\,X(\tau, \sigma + 2\pi)~=~g\,h\,X(\tau, \sigma)\;,
\end{equation}
i.e., the constructing element $g$ is invariant under the action of
$h$,
\begin{equation}
\label{eqn:constr_element_case1} |g\rangle  \
\stackrel{h}{\rightarrow} \ |h\,g\,h^{-1}\rangle~=~|g\rangle \;.
\end{equation}
$hX$ closes under the same constructing element $g$ as $X$. Thus,
both give rise to the same Hilbert space ${\mathcal H}_g \
\stackrel{h}{\rightarrow} \ {\mathcal H}_{hgh^{-1}} = {\mathcal
H}_g$. Furthermore, on the orbifold space ${\mathbb R}^6/S$ the
string coordinates $hX$ and $X$ are identified. Thus, $hX$ and $X$
describe the same physical state.

In summary, provided a constructing element $g$, we have shown that
for commuting elements $h$, $h\,X$ and $X$ give rise to the same
physical state from the same Hilbert space. Since $h$ has to act as
the identity on $|\text{phys}\rangle$, the following condition
follows using Eqs.~\eqref{eqn:physicalstate1},
\eqref{eqn:trafophase} and \eqref{eqn:constr_element_case1}:
\begin{equation}
q_\mathrm{sh}\cdot v_h - P_\mathrm{sh}\cdot V_h ~\stackrel{!}{=}~0
\mod 1\;.
\end{equation}

\subsubsection*{Non--commuting elements: $\boldsymbol{[h,g] \neq 0}$}

Next, considering a non--commuting element $h$ in
Eq.~\eqref{eqn:boundarytimesh} yields
\begin{equation}
h\,X(\tau, \sigma + 2\pi) ~=~ \left(h\,g\,h^{-1}\right)\,h\, X(\tau,
\sigma)\;,
\end{equation}
i.e., the constructing element $g$ is not invariant under the action
of $h$,
\begin{equation}
\label{eqn:mapgtohgh} |g\rangle  ~ \xrightarrow{h}~
|h\,g\,h^{-1}\rangle \neq |g\rangle\;.
\end{equation}
In the upstairs picture, i.e. in the covering space ${\mathbbm R}^6$
of the orbifold ${\mathbbm R}^6/S$, one has different Hilbert spaces
for the states with boundary conditions $g$ and $h\,g\,h^{-1}$. In
this picture, Eq.~\eqref{eqn:mapgtohgh} says that $h$ maps states
from a given Hilbert space $\mathcal{H}_g$ onto a different Hilbert
space $\mathcal{H}_{h\,g\,h^{-1}}$. Subsequent application of $h$
then leads to the sequence \footnote{Note that in all ${\mathcal
H}_{h^ngh^{-n}}$ the left--moving momenta $P_\mathrm{sh}$ of
equivalent states are identical. The same holds for
$q_\mathrm{sh}$.}
\begin{equation}
\mathcal{H}_g ~ \xrightarrow{h}~ \mathcal{H}_{h\,g\,h^{-1}}~
\xrightarrow{h}~ \mathcal{H}_{h^2\,g\,h^{-2}}~ \xrightarrow{h}~
\mathcal{H}_{h^3\,g\,h^{-3}} ~ \xrightarrow{h}~ \ldots\;.
\end{equation}
The crucial point is now that on the orbifold $h\,X$ and $X$ are
identified. This means that, on the orbifold, the different Hilbert
spaces $\mathcal{H}_{h^n\,g\,h^{-n}}$ of the upstairs picture are to
be combined into a single orbifold Hilbert space. Invariant states
are then linear combinations of states from all
$\mathcal{H}_{h^n\,g\,h^{-n}}$. Such linear combinations do, in
general, involve relative phase factors (often called
\emph{gamma--phase} $\gamma$). So, the new ansatz for a physical
state reads:
\begin{eqnarray}
|\text{phys}\rangle & \sim & \sum_{n} \left( e^{-2\pi \I\, n\,
\gamma}\, |q_\mathrm{sh}\rangle_\mathrm{R} \otimes
|P_\mathrm{sh}\rangle_\mathrm{L} \otimes
|h^n\,g\,h^{-n}\rangle\right)
\nonumber\\
    &   =  & |q_\mathrm{sh}\rangle_\mathrm{R} \otimes |P_\mathrm{sh}\rangle_\mathrm{L}
    \otimes \left(\sum_{n} e^{-2\pi \I\, n\, \gamma}\ |h^n\,g\,h^{-n}\rangle\right)\;,
\label{eqn:physicalstate2}
\end{eqnarray}
where $\gamma = \text{integer}/N$, $N$ being the order of the
orbifold. The geometrical part of the linear combination transforms
non--trivially under $h$
\begin{equation}
\label{eqn:constr_element_case2} \sum_{n} e^{-2\pi \I n \gamma}\
|h^n\,g\,h^{-n}\rangle\, \stackrel{h}{\rightarrow} \ e^{2\pi \I\,
\gamma} \sum_{n} e^{-2\pi \I\, n\, \gamma}\ |h^n\,g\,h^{-n}\rangle
\;.
\end{equation}
Since $h$ has to act as the identity on $|\text{phys}\rangle$, the
following condition follows using Eqs.~\eqref{eqn:trafophase},
\eqref{eqn:physicalstate2} and \eqref{eqn:constr_element_case2} for
non--commuting elements:
\begin{equation}
\label{eqn:projectionwithgamma} q_\mathrm{sh}\cdot v_h -
P_\mathrm{sh}\cdot V_h + \gamma~\stackrel{!}{=}~0 \mod 1\;.
\end{equation}
Notice that $\gamma$ depends on $h$. Thus we can always choose
$\gamma (h)$ such that this condition is satisfied\footnote{In this
sense, building linear combinations and computing the $\gamma$ phase
is not a projection condition. Note that $\gamma (h)$ is
well--defined: if $h_1gh_1^{-1} = h_2gh_2^{-1}$ then $\gamma (h_1) =
\gamma (h_2)$.}. In principle, these steps have to be repeated for
all non--commuting elements in order to ensure invariance of the
physical state under the action of the whole orbifold group $O
\subset S \otimes G$. The result for $|\text{phys}\rangle$ reads
\begin{equation}
|\text{phys}\rangle ~=~ |q_\mathrm{sh}\rangle_\mathrm{R} \otimes
|P_\mathrm{sh}\rangle_\mathrm{L} \otimes \left( \sum_{h =
\mathbbm{1} \text{ or } [h,g]\neq 0} e^{-2\pi \I \gamma(h)}\
|h\,g\,h^{-1}\rangle \right)\;,
\end{equation}
where the summation over $h$ is such that each term
$|h\,g\,h^{-1}\rangle$ appers only once. Note that the summation
over $h$ can be understood as a summation over all elements of the
conjugacy class of $g$.

\subsubsection*{Example}

\begin{figure}[!t]
\centerline{\input{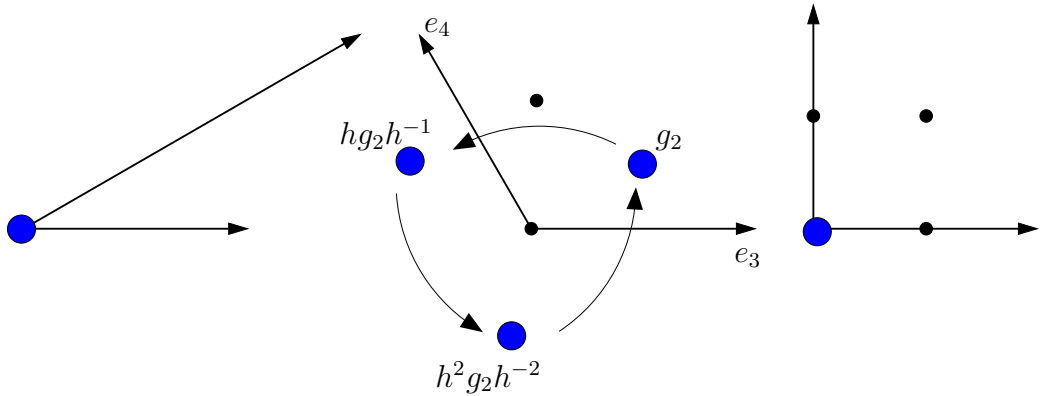}} \caption{Illustration
of the $\gamma$--factor.  The fixed point associated with  the space
group element $g_2=(\theta,\,e_3)$ is invariant under
$(\theta,\,e_3)$, but transforms into equivalent fixed points
outside the fundamental domain under $h=(\theta,0)$. To form an
eigenstate of $(\theta,0)$, one needs to build linear combinations
of  the equivalent fixed points. The corresponding eigenvalues can
be $1, \mathrm{e}^{\pm 2\pi \I/3}$.} \label{fig:gammarule}
\end{figure}

To illustrate the construction of physical states, let us consider
an example in the first twisted sector of the $\mathbb{Z}_6$--II
orbifold. In the SU(3) lattice spanned by $e_3$ and $e_4$, there are
three inequivalent fixed points associated to the constructing
elements $g_1=(\theta,\, 0)$, $g_2=(\theta,\, e_3)$ and
$g_3=(\theta,\, e_3+e_4)$, or analogously $g_i = (\theta,\, a_i\,
e_3 + b_i\, e_4)$ for $i=1,2,3$ with $a_i=(0,\,1,\,1)$ and
$b_i=(0,\,0,\,1)$. Then, restricting to the SU(3) lattice, the
geometrical part of a physical state can be written as
\begin{equation}
\label{eqn:SU3_linear_combination} \sum_{n,m} e^{-2\pi \I (n+m)
\gamma}\ \big|(\theta,(n+m+a_i)\,e_3+(2m-n+b_i)\,e_4)\big\rangle \;.
\end{equation}
Since the action of $\theta$ in the SU(3) lattice has order 3, the
only possible $\theta$--eigenvalues of
Eq.~\eqref{eqn:SU3_linear_combination} have
$\gamma=0,\,\pm\frac{1}{3}$. In the case of $\gamma=0$,
Eq.~\eqref{eqn:SU3_linear_combination} is invariant under all
rotations and translations for all three $g_i$. However, if
$\gamma=\pm\frac{1}{3}$,  the eigenvalue of
Eq.~\eqref{eqn:SU3_linear_combination} depends on $g_i$: for the
fixed point at the origin associated to $g_1$,
Eq.~\eqref{eqn:SU3_linear_combination} is invariant under $\theta$,
but has an eigenvalue $\mathrm{e}^{2 \pi \I\,\gamma\,(k+l)}$ under
$(\mathbbm{1}, ke_3+le_4)$. Similarly, for the fixed points away
from the origin, corresponding to $g_i$ ($i \neq 1$),
Eq.~\eqref{eqn:SU3_linear_combination} picks up a phase
$\mathrm{e}^{- 2 \pi \I\,\gamma\,(a_i + b_i)}$ under $\theta$ (see
Fig.~\ref{fig:gammarule}).  It can be shown that for physical states
$\gamma \neq 0$ is only possible in the presence of a Wilson line in
the $e_3$ and $e_4$ directions.

\subsection{String selection rules}
\label{sec:ourstringselectionrules}

Consider the $n$--point correlation function of two fermions and
$n-2$ bosons~\cite{Dixon:1986qv, Hamidi:1986vh}
\begin{equation}
\label{eqn:correlation} \langle
\text{F\,F\,B}\ldots\text{B}\rangle\;.
\end{equation}
The corresponding physical states shall be denoted by $\Psi_i$, $i =
1,\ldots,n$. Then, in the field theory limit, a non--vanishing
correlation function induces the following term in the
superpotential
\begin{equation}
W~\supset~\Psi_1\, \Psi_2\, \Psi_3 \ldots \Psi_n\;.
\end{equation}
A complete evaluation of Eq.~\eqref{eqn:correlation} has only been
performed for 3--point couplings and yields a moduli dependent
coupling strength~[\citen{Dixon:1986qv,Hamidi:1986vh},%
\citen{Casas:1991ac,Erler:1992gt}].

On the other hand, symmetries of Eq.~\eqref{eqn:correlation} give
rise to the so--called string selection rules. These rules determine
whether a given coupling vanishes or not. We use the following
notation: the constructing elements of $\Psi_i$ are denoted by $g_i
\in S$ and their left- and right--moving shifted momenta, by
$P_{\mathrm{sh},i}$ and $q_{\mathrm{sh},i}$, respectively. Then, the
string selection rules read:

\begin{enumerate}
\item {\bf Gauge invariance}

The sum over all left--moving shifted momenta $P_{\mathrm{sh},i}$
must vanish:
\begin{equation}
\label{eqn:gaugeinvariance} \sum_i P_{\mathrm{sh},i} = 0
\end{equation}
This translates to the field theoretic requirement of gauge
invariance for allowed terms in the superpotential.

\item {\bf Conservation of R--charge}

R--charge is defined by
\begin{equation}
R_i^a~=~q_{\mathrm{sh},i}^a - N_i^a + N_i^{*a} \quad\text{ for
}a=0,\ldots,3\;,
\end{equation}
where $N_i^a$ and $N_i^{*a}$ are integer oscillator numbers,
counting the number of excitations with oscillators
$\tilde{\alpha}^a$ and $\tilde{\alpha}^{\overline a}$, respectively.
Then the conditions~\cite{Kobayashi:2004ya}
\begin{equation}
\label{eqn:rchargeconservation} \sum_i R_i^a~=~0 \mod  N^a
\quad\text{ for }a=1,2,3
\end{equation}
have to be imposed, where $N^a$ denotes the order of the twist
component $v^a$ in the $a$--th complex plane, i.e.\ $N^a v^a \in
\mathbb{Z}$ (no summation). Here, two of the $R_i$ come from
fermions and the rest from bosons. For computational purposes, it is
more convenient to use the purely bosonic notation, where
Eq.~\eqref{eqn:rchargeconservation} becomes $\sum_i R_i^a = -1
\text{ mod } N^a$.

This condition can be understood as a remnant of 10 dimensional
Lorentz invariance.

\item {\bf Space group selection rule}

The product of constructing elements $g_i$ must be the identity:
\begin{equation}
\prod_i g_i = (\mathbbm{1}, 0) \ .
\end{equation}
In terms of conjugate elements $h_i g_i h_i^{-1}$ of $g_i$, this
condition can be reformulated as $\prod_i h_ig_ih_i^{-1} =
(\mathbbm{1}, v)$ with $v \in \sum_i(\mathbbm{1} -
\theta^{k_i})\Lambda$~\cite{Kobayashi:1990mc}.

This selection rule can be visualized as the geometrical ability of
twisted strings to join.
\end{enumerate}

\subsection{On the need for a $\boldsymbol{\gamma}$ selection rule}
In the literature, there exists an additional selection rule, here
referred to as the $\gamma$ rule. In our notation, it
reads~\cite{Casas:1991ac, Kobayashi:2004ya}
\begin{equation}
\sum_i \gamma_i = 0 \text{ mod }1\;,
\end{equation}
where $\gamma_i$ denotes the gamma--phase of $\Psi_i$. In this
section, we argue that, in contrast to previous statements, a fully
consistent approach yields to automatic fulfillment of the $\gamma$
rule.

The correlation function corresponding to the coupling
\begin{equation}
\label{eqn:couplingpsi}
 \Psi_1\, \Psi_2 \dots \Psi_n
\end{equation}
should be invariant under the action of the full space group. Let us
assume first that the states $\Psi_i$ corresponded to linear
combinations of equivalent fixed points within the fundamental
domain of the torus (see e.g.~\cite{Casas:1991ac, Kobayashi:1991rp,
Kobayashi:2004ya}). For example, in the case of the
$\mathbb{Z}_6$--II orbifold only fixed points in the $G_2$ lattice
could form linear combinations. Under this assumption, different
states $\Psi_i$ would be eigenstates with respect to
\emph{different} space group elements. So one could not transform
the coupling Eq.~\eqref{eqn:couplingpsi} with a given $h =
(\theta^l,m_\alpha e_\alpha)$. Thus the fully consistent approach
for building invariant linear combinations, as presented in
Appendix~\ref{sec:physstates}, is necessary. In this case, we can
compute the gamma--phase for all states $\Psi_i$ from
Eq.~\eqref{eqn:projectionwithgamma}, i.e. $\gamma_i = \gamma_i (h)$
for arbitrary $h=(\theta^l,m_\alpha e_\alpha)$. But since allowed
couplings already fulfill the selection rules
Eqs.~\eqref{eqn:gaugeinvariance} and
\eqref{eqn:rchargeconservation}, the $\gamma$ rule is satisfied
trivially\footnote{Also in the presence of oscillators, the $\gamma$
rule is satisfied automatically.}:
\begin{eqnarray}
\gamma_i(h)                    & = & P_{\mathrm{sh},i} \cdot V_h -
q_{\mathrm{sh},i} \cdot v_h
\;, \\
\Rightarrow \sum_i \gamma_i(h) & = & \underbrace{\left(\sum_i
P_{\mathrm{sh},i}\right)}_{=0 \text{ see
Eq.~\eqref{eqn:gaugeinvariance}}}\cdot V_h -
\underbrace{\left(\sum_i q_{\mathrm{sh},i}\right)}_{\sim 0 \text{
see
Eq.~\eqref{eqn:rchargeconservation}}}\cdot v_h\;, \\
& = & 0 \mod  1\;.\nonumber
\end{eqnarray}
Thus, the $\gamma$ rule in the fully consistent approach is
\emph{not} a selection rule. It is a consequence of other selection
rules and invariance of the states. We therefore conclude that the
coupling must only satisfy gauge invariance, $R$--charge
conservation and the space group selection rule.

This has important consequences. For example, in the model A1
of~\cite{Kobayashi:2004ya}, there is no mass term for the exotics
$\bar{q}_2 q_2$ up to order 9 in singlets. However, we find that the
coupling $\bar{q}_2 q_2 S_9 S_{15} S_{22} S_{33}$ is allowed by the
selection rules of Appendix~\ref{sec:ourstringselectionrules}.
Further, using the prescription of Appendix~\ref{sec:physstates},
the gamma--phases of the corresponding physical states are $\gamma_i
= (\tfrac{1}{2},\,0,\,0,\,\tfrac{5}{6},\,\tfrac{2}{3},\,0)$ for $h =
(\theta,0)$ , which sum up to $2$. This is in contrast
to~\cite{Kobayashi:2004ya}, where $\gamma_i =
(0,\,0,\,0,\,\tfrac{1}{2},\,\tfrac{2}{3},\,0)$ and linear
combinations were built differently.

\section{$\boldsymbol{D}$--flatness} \label{sec:Dflat}

In this appendix, a simple method is reviewed that allows to analyze
$D$--flatness. It also provides a simple test whether it is
possible to cancel the FI term with a given set of fields.

Let us start by briefly reviewing the issue of $D$-flatness and
cancellation of the FI term
\cite{Buccella:1982nx,Gatto:1986bt,Font:1988tp,Dine:1987xk,Cleaver:1997jb}. In
supersymmetric theories, there is the so-called $D$-term potential.
In the case of a \U1 gauge theory it is given by
\begin{equation}\label{eq:VD}
 V_D~\propto~\left[\sum_i q_i\,|\phi_i|^2\right]^2\;.
\end{equation}

Consider as a first example a \U1 gauge theory with two fields
$\phi_\pm$ carrying the charges $\pm1$. Clearly, as long as
$|\phi_+|=|\phi_-|$, $V_D$ vanishes. That is, one has a $D$-flat
direction, parametrized by $x=|\phi_+|=|\phi_-|$.

Consider now a theory with one field ($\phi_1$) with charge $2$ and
two fields ($\phi_2,\phi_3$) with charges $-1$.\footnote{For the
moment, we ignore anomalies.}  Then we have many flat directions,
described by the roots of the equation $2 |\phi_1|^2 - |\phi_2|^2 -
|\phi_3|^2 = 0$.  It is convenient to associate these directions to
the (holomorphic) monomials
\[ \phi_1\,\phi_2^2\;,\quad
\phi_1\,\phi_3^2\;,\quad \phi_1\,\phi_2\,\phi_3\;,\]
respectively.
That is, a monomial $\phi_{1}^{n_1}\,
\phi_{2}^{n_2}\cdots\phi_k^{n_k}$ represents a flat direction,
defined by the relation
\[
 \frac{|\phi_1|}{\sqrt{n_1}}~=~\frac{|\phi_2|}{\sqrt{n_2}}~=~\dots
 ~=~\frac{|\phi_k|}{\sqrt{n_k}}
 \quad\text{and}\quad|\phi_j|~=~0~~\text{for}~~n_j~=~0\;.
\]
The crucial feature of such monomials is that they are (obviously)
gauge invariant. More precisely, every holomorphic gauge invariant
monomial represents a $D$-flat direction \cite{Buccella:1982nx}.

It is, however, clear that there is only a finite number of linearly
independent $D$-flat directions. In the previous example, the third
direction is not independent of the other two. In other words, the
requirement $V_D=0$ poses only one constraint on the three real
variables ($|\phi_i|^2$) entering (\ref{eq:VD}). The space of
absolute values $|\phi_i|$ is 2-dimensional. The power of using the
monomials is that checking whether certain monomials are linearly
independent or not is fairly simple: identify with each monomial the
vector of exponents, $v=(n_1,n_2\dots )$. The directions are
independent if and only if the vectors are linearly independent. In
the previous example one would get the vectors $(1,2,0)$, $(1,0,2)$,
and $(1,1,1)$, out of which only two are linearly independent.

It is also clear how to obtain these vectors: all of them are orthogonal to the
vector of charges $q=(q_1, q_2\dots)$. That is, the problem of finding the above
monomials (and thus the $D$-flat directions) is reduced to the problem of
finding vectors $v$ with the following properties: \begin{enumerate}
 \item $q\cdot v=0$,
 \item $v_i\in \mathbbm{N}_0$.
\end{enumerate}
The property that the $v_i$ be integer-valued does not pose a
constraint in our models: since the charges are rational,
one can rescale any $v$ having the first property such as to have
integer entries. However, the requirement that the entries be
non-negative, which reflects that the monomials ought to be
holomorphic, is a constraint.

The discussion so far can easily be extended to $\U1^n$ theories.
Here the $D$-term potential is
\begin{equation}
 V_D~\propto~\sum\limits_{j=1}^n\left[\sum_i q_i^{(j)}\,|\phi_i|^2\right]^2\;,
\end{equation}
where $q_i^{(j)}$ is the  charge of the field $\phi_i$ under the
$j^\mathrm{th}$ \U1 factor. Now a $D$-flat direction has to satisfy
the above constraints for each \U1 factor separately. Again, it is
advantageous to represent $D$-flat directions by holomorphic gauge
invariant monomials (dubbed `HIMs' in the literature
\cite{Cleaver:1997jb}). Then the vector $v$ of exponents has to be
orthogonal to every charge vector
$q^{(j)}=(q_1^{(j)},q_2^{(j)},\dots)$. In other words, $v$ has to be
in the kernel of the charge matrix $Q$,
\begin{equation}
 Q\cdot v~=~0\;,\quad\text{with}\quad
 Q~=~\left(\begin{array}{ccc}
 q_1^{(1)} & q_2^{(1)} & \dots\\
 q_1^{(2)} & q_2^{(2)} & \dots\\
 \vdots & \vdots & \vdots\\
 q_1^{(n)} & q_2^{(n)} & \dots\\
 \end{array}\right)\;.
\end{equation}
Hence, the problem of finding the $D$-flat directions of a $\U1^n$
gauge theory is reduced to the task of calculating the kernel of the
charge matrix $Q$, and to forming linear combinations of elements of
this kernel in such a way that the entries are non-negative
integers. The maximal linear independent set of such linear
combinations is in one-to-one correspondence with the independent
$D$-flat directions.

Next, let us comment on what happens if there are non-Abelian gauge
factors. Then the $D$-term potential is to be amended by
\begin{equation}
 V_D^\mathrm{non-Abelian}
 ~\propto~
 \sum_a\left[\sum_i \phi_i^\dagger\,\mathsf{T_a}\, \phi_i\right]^2 \
 , \end{equation}
with $\mathsf{T}_a$ denoting the group generators. It is
straightforward to see that the results obtained so far generalize
to the non-Abelian case \cite{Buccella:1982nx}: the $D$-flat
directions are again in correspondence with holomorphic gauge
invariant monomials. That is, one can amend the monomials discussed
so far such as to include fields transforming non-trivially under
non-Abelian gauge factors, as long as these fields are contracted in
such a way that the monomials are gauge invariant.

Finally, let us review the issue of cancelling the FI term. For an
`anomalous' \U1, the $D$-term potential \eqref{eq:VD} gets modified
to
\begin{equation}
 V_D^\mathrm{anom}
 ~\propto~
 \left[\sum\limits q_i^\mathrm{anom}\,|\phi_i|^2+\xi\right]^2\;,
\end{equation}
where in our convention $\xi>0$. To cancel the FI term one thus has
to find a holomorphic monomial,
\begin{equation}
 I~=~\phi_1^{n_1}\,\phi_2^{n_2}\,\dots
\end{equation}
with net negative charge under $\U1_\mathrm{anom}$, i.e.\
\begin{equation}
 \sum_i n_i\,q_i^\mathrm{anom}~<~0\;.
\end{equation}

To summarize, the $D$-flat directions are in one-to-one
correspondence with holomorphic gauge invariant monomials. In the
Abelian case, such monomials can be identified with elements of the
kernel of the charge matrix $Q$ with non-negative integer entries.
Cancelation of the FI term requires the existence of a holomorphic
monomial with net negative charge under $\U1_\mathrm{anom}$, which
is gauge invariant with respect to all other group factors.

\section{Family reflection symmetry and Matter Parity,
$\boldsymbol{\mathbb{Z}_2^{\mathcal{M}}}$}

We would like to define an effective low energy theory which
preserves  R parity.  This has the advantage of greatly reducing the
number of arbitrary parameters in the superpotential,  forbidding
dimension 3 and 4 baryon or lepton number violating operators, and
preserving a viable dark matter candidate, i.e. the LSP. Our
strategy for accomplishing this is, in principle, quite simple.  We
make use of ``family reflection symmetry" or ``matter parity"
defined as a discrete subgroup of $\U1_{\rm B-L}$. This is a global
$\mathbb{Z}_2^{\cal M}$ symmetry (commuting with supersymmetry)
which is \emph{even} on the Higgs doublets and \emph{odd} on all SM
quark and lepton fields. It forbids the following dangerous baryon
or lepton number violating operators,
\begin{equation}  \overline{u}\,\overline{d}\,\overline{d}\;,\quad
q\,\overline{d}\, \ell\;,\quad \ell\, \ell\,\overline{e}
\quad\text{and} \quad \ell\, h_u\;.  \label{eq:blviol}
\end{equation}
On the other hand, it allows quark and lepton Yukawa couplings as well as
the Majorana neutrino mass operator $\bar \nu \bar \nu$.

Consider the effective operators
\begin{eqnarray}
& \mathcal{O}\,\langle s_1 \dots s_n \rangle \;\; {\rm with} \;\;
\mathcal{O} \;\; {\rm in} \;\; (\ref{eq:blviol}) \ , & \label{eq:O}  \\
& \lefteqn{\overline{\nu} \; \overline{\nu} \,  \langle  s_1' \dots
s_n' \rangle } \  \label{eq:nunu}  & .
\end{eqnarray}
We want to forbid the dangerous proton decay operators (\ref{eq:O}),
while allowing for Majorana neutrino masses (\ref{eq:nunu}). This
puts a constraint on the $B-L$ charges of the SM singlets which get
non--zero VEVs. In particular, it requires
\begin{equation}
 -1 + \sum_i q_i ~ \not=  ~ 0
\end{equation}
for any set of singlets with non--zero VEVs, where $q_i$
are the $B-L$ charges. In addition,
\begin{equation}
 2 +  \sum_i q_i' ~=~ 0
\end{equation}
must be satisfied for at least one singlet configuration with $B-L$ charges
$q'_i$. In our theory, the singlets come in pairs with opposite $B-L$ charges
and these charges are rational. Then the relevant solution to the above
equations is
\begin{equation}\label{eq:q_i}
q_i~=~  \pm {   2 k_i   \over  2l_i +1  }
\end{equation}
for integer $k_i$ and $l_i$, with the additional condition that
adding/subtracting the numerators of $q_i$ can yield 2, i.e.\
\begin{equation}
\sum_i k_i\, N_i~=~1 \;,  \label{eq:k}
\end{equation}
for some $N_i \in \mathbbm{Z}$.  For example, the numerators can differ by 2.
If  there are only two fields with charges $\pm q_a$, the corresponding
constraint is  $ q_a = \pm  2/(2l+1)$.

The above singlet VEVs break $\U1_{B-L}$ to a discrete subgroup.
Consider an element of $\U1_{B-L}$ defined by
\begin{equation}
\widetilde{R}(\alpha)~=~e^{\I\, \pi\, \alpha \,\mathsf{t}_{B-L}}\; .
\end{equation}
The $B-L$ and $\tilde R(\alpha)$ charges of SM particles are given in
Table \ref{tab:BminusL}.  The choice $\alpha = 3$ corresponds
to family reflection symmetry (FRS).
 If  only the singlets satisfying
\begin{equation}
q_{B-L}(\widetilde{s})~=~\pm \frac{2}{3} \mathbb{Z} 
\end{equation}
obtain VEVs and there is at least one singlet for which $\mathbb{Z}
\neq 0$, then $\U1_{B-L}$ is broken to $\widetilde{R}(3) \equiv $
FRS.
Clearly, products of these singlets can contribute to Yukawa couplings for
quarks and leptons. Further, products of singlets with $B-L$ charge $-2/3$ (or,
more generally,  those with   $B-L$ charge $2/3$ and $-4/3$, etc.)  can
generate   Majorana neutrino masses. On the other hand, the proton decay
operators are forbidden.

\begin{table*}[t!]
\centerline{
\begin{tabular}{|c||c|c||c|c||c|c|c|c|}
\hline
  & quarks & leptons & $h_u$ & $h_d$ &  $\bar u \bar d \bar d$ &  $q \bar d \ell$ &  $\ell \ell \bar
  e$ &  $\ell h_u$  \\
\hline
&&&&&&&& \\
B-L   &1/3 &-1 &0 &0 & -1 & -1  &  -1 &  -1  \\
\hline &&&&&&&& \\ $\tilde R$ & $e^{\I\, \alpha\, \pi/3}$ & $e^{- \I\,
\alpha\,
\pi}$ & 1 & 1 & $e^{-\I\, \alpha\, \pi}$ & $e^{- \I\, \alpha\, \pi}$ & $e^{- \I\,
\alpha\, \pi}$  &  $e^{- \I\, \alpha\, \pi}$ \\
\hline
\end{tabular}
}

\caption{$B-L$ and $\tilde R$ charges for SM particles, with
opposite $B-L$  charges for anti-particles, and for baryon and
lepton number violating operators with dimension $\leq 4$.
\label{tab:BminusL} }
\end{table*}

It is  possible to generalize FRS to a $\mathbb{Z}_N$ group.  In
general as long as $\alpha \neq 2 \mathbb{Z}$,  $\tilde R(\alpha)$
will forbid the dangerous operators, Eq.~\eqref{eq:blviol}, and
allow all Yukawa couplings. Consider a field $\phi \subset
\{\widetilde{s}\}$ with $ q_{B-L}(\phi) = f \ $.  Such a field
breaks $\U1_{B-L}$ to the subgroup $\tilde R(\alpha)$ with $\alpha =
2/f$. The effective Majorana neutrino mass operator  $\bar \nu \bar
\nu \phi^n $ is allowed for  $[2+ n f]/f = \mathbb{Z}$  or  $\alpha
\equiv 2/f = \mathbb{Z}$. Hence, for $odd$ $\alpha$, we can  both
forbid the dangerous operators, Eq.~\eqref{eq:blviol} and  obtain
non--zero Majorana neutrino mass. The corresponding constraint on
$f$ is then  $ f = \pm 2/(2 \mathbb{Z} + 1)$, as expected.  For
example,
\begin{enumerate}
\item $f=\pm 2, \; \alpha = 1$  gives  $\widetilde{R} = e^{\I\, \pi\, q_{B-L}} \in
\mathbb{Z}_6$,
\item $f=\pm 2/3, \; \alpha = 3$  gives  $\widetilde{R} = e^{3 \,\I\,\pi\, q_{B-L}} \in
\mathbb{Z}_2$,
\item $f=\pm 2/5, \; \alpha = 5$  gives  $\widetilde{R} = e^{5 \,\I\,\pi\, q_{B-L}} \in
\mathbb{Z}_6$,
\item $f=\pm 2/7, \; \alpha = 7$  gives  $\widetilde{R} = e^{7 \,\I\,\pi\, q_{B-L}} \in
\mathbb{Z}_6$,
\item $f=\pm 2/9, \; \alpha = 9$  gives  $\widetilde{R} = e^{9 \,\I\,\pi\, q_{B-L}} \in
\mathbb{Z}_2$.
\end{enumerate}

This is easily generalized to configurations with many different
singlets getting VEVs (with the constraints given in \eqref{eq:q_i}
and \eqref{eq:k}). These conserve  matter parity,
$\mathbb{Z}_2^{\cal M}$.

\section{Search for $\boldsymbol{B-L}$ and R parity}

Our search for $\U1_{B-L}$ is based on the methods developed in
ref.~\cite{Raby:2007yc} (for an earlier discussion of $\U1_{B-L}$
and its applications see \cite{Buchmuller:2006ik} and
\cite{Lebedev:2006kn}). In Tab.~\ref{tab:spectrum_SM}, we list the
standard model particle content with their hypercharge and $B-L$
charges.

\begin{table}[h!]
\centering
\normalsize
\renewcommand{\arraystretch}{1.4}
\begin{tabular}{|c|l||c|l||c|l|}
\hline
$q$ & $(\boldsymbol{3}, \boldsymbol{2})_{\phantom{\text{-}}1/6, \phantom{\text{-}}1/3}$       &   $\ell$ & $(\boldsymbol{1}, \boldsymbol{2})_{\text{-}1/2, \text{-}1}$   &  $h_u$ & $(\boldsymbol{1}, \boldsymbol{2})_{\phantom{\text{-}}1/2, \phantom{\text{-}}0}$     \\
$\bar{u}$ & $(\boldsymbol{\overline{3}}, \boldsymbol{1})_{\text{-}2/3, \text{-}1/3}$ &   $\bar{e}$ & $(\boldsymbol{1}, \boldsymbol{1})_{\phantom{\text{-}}1, \phantom{\text{-}}1}$    &  $h_d$ & $(\boldsymbol{1}, \boldsymbol{2})_{\text{-}1/2, \phantom{\text{-}}0}$\\
\cline{5-6}
$\bar{d}$ & $(\boldsymbol{\overline{3}}, \boldsymbol{1})_{\phantom{\text{-}}1/3, \text{-}1/3}$    &   $\bar{\nu}$ & $(\boldsymbol{1}, \boldsymbol{1})_{\phantom{\text{-}}0, \phantom{\text{-}}1}$    &  \multicolumn{2}{|c|}{}            \\
\hline
\end{tabular}
\caption{Matter content of the standard model, where the subscripts
denote hypercharge and $B-L$, respectively. In our conventions,
${\displaystyle Q=T_{3L} + Y}$.} \label{tab:spectrum_SM}
\end{table}

The choice for $\mathrm{U}(1)_{B-L}$ depends on the choice of
hypercharge in the first place. In this publication, we do not take
the most general approach, but assume that hypercharge is given by
$\mathrm{SO}(10) \supset \mathrm{SU}(5) \supset \mathrm{SU}(3)
\times \mathrm{SU}(2) \times \mathrm{U}(1)_Y$. Furthermore, we
demand that the first and second families come from
$\boldsymbol{16}$-plets localized in the first twisted sector,
whereas the multiplets of the third family may come from any sector
of the theory.

To find a suitable $\mathrm{U}(1)_{B-L}$, we proceed as follows. In
general, the shift and Wilson lines break the gauge group in 10 dimensions
\begin{equation}
\mathrm{E}_{8}\times\mathrm{E}_{8}' \quad \rightarrow \quad \text{non-Abelian} \times \mathrm{U}(1)^n
\;.
\end{equation}
The $\mathrm{U}(1)$ generators are $n$ linearly independent
directions $\mathsf{t}_i$ in the root lattice of
$\mathrm{E}_{8}\times\mathrm{E}_{8}'$ that are orthogonal to the
simple roots of the unbroken non-abelian gauge group. For the $B-L$
direction, we make the general ansatz
\begin{equation}
\mathsf{t}_{B-L}~ =~x_1\, \mathsf{t}_1 + x_2\, \mathsf{t}_2 + \ldots
+  x_n\, \mathsf{t}_n\;.
\end{equation}
The $B-L$ charge of a particular representation is given by the scalar product
of its highest weight and $B-L$. We denote the highest weights of the
left-handed quark doublets and of the right-handed quark singlets by
$\Lambda_i$, $i=1,\ldots,9$. Note that the first two families are fixed, and we
loop over all representations which have the right quantum numbers to be the
quarks of the third generation. For each such choice, we have
\begin{equation}
\Lambda_i \cdot \mathsf{t}_{B-L}~=~x_1 \, \Lambda_i \cdot \mathsf{t}_1
+ x_2 \, \Lambda_i \cdot \mathsf{t}_2 + \ldots +  x_n \,
\Lambda_i \cdot \mathsf{t}_n\;,
\end{equation}
which is a system of 9 linear equations. Although one may think that these 9
equations severely constrain the values of $x_i$, this is not true. In general,
the system will be under-determined, since the quarks may differ by
localization, but not necessarily by the highest weights of their gauge
representations. In order to account for the $B-L$ charges of the
leptons and Higgses, and for the absence of chiral exotics, we set up necessary,
but in general \emph{not sufficient}, linear constraints:
\begin{align}
\sum_{(\boldsymbol{3},\boldsymbol{2}),(\boldsymbol{\overline{3}},\boldsymbol{2})}
q_{B-L}&=~1\;,& \quad
\quad \sum_{(\boldsymbol{3},\boldsymbol{1}),(\boldsymbol{\overline{3}},\boldsymbol{1})}
q_{B-L}&=~-2\;,\nonumber\\
\sum_{(\boldsymbol{1},\boldsymbol{2})} q_{B-L}&=~-3\;, &
\sum_{(\boldsymbol{1},\boldsymbol{1}), \,\,Y\neq0} q_{B-L}&=~3\;.
\label{eq:linear_constraint}
\end{align}
For readability, we use $q_{B-L}$ in the above equations as a shorthand for
$\Lambda_k \cdot \mathsf{t}_{B-L}$, where $\Lambda_k$ runs over the highest
weights of the representations in the sum. Note that the sum over e.g.\ the
$(\boldsymbol{3},\boldsymbol{1})$ and
$(\boldsymbol{\overline{3}},\boldsymbol{1})$ representations reduces to that
over the right-handed quarks $\bar{u}$ and $\bar{d}$ alone, since we assume
pairs of exotic particles to carry $B-L$ charge assignments that are
equal in magnitude but opposite in sign. Another linear constraint comes from
the requirement that $\mathrm{U}(1)_{B-L}$ be non-anomalous. If the
model has an anomalous $\text{U}(1)$ direction that we will denote by
$\mathsf{t}_\mathrm{anom}$, we demand that it be orthogonal to $B-L$:
\begin{equation}
\left(x_1\,\mathsf{t}_1 + x_2\,\mathsf{t}_2 + \ldots +  x_n\,\mathsf{t}_n\right)
\cdot \mathsf{t}_\mathrm{anom}~=~0\;.
\label{eq:linear_anomaly_constraint}
\end{equation}
In principle, we could write down many more linear conditions,
e.g.~we could demand that the sum of the $B-L$ charges of $u$-type
quarks like
$(\boldsymbol{\overline{3}},\boldsymbol{1})_{{\text{-}2/3},{\text{-}1/3}}$
and $(\boldsymbol{3},\boldsymbol{1})_{{\text 2/3},{\text 1/3}}$
alone gives -1, which is more constraining than
Eq.~(\ref{eq:linear_constraint}). However, our experience shows that
it is more practical to drop these conditions, since the non-linear
equations to be introduced below are constraining enough.


The linear equations by no means exhaust the constraints we may require to be
fulfilled by a vector-like spectrum. In particular, there are the cubic,
\begin{align}
\sum_{(\boldsymbol{3},\boldsymbol{2}),(\boldsymbol{\overline{3}},\boldsymbol{2})}
(q_{B-L})^3& =~\frac{1}{9}\;, &
\quad
\sum_{(\boldsymbol{3},\boldsymbol{1}),(\boldsymbol{\overline{3}},\boldsymbol{1})}
(q_{B-L})^3&=~-\frac{2}{9}\;, \nonumber\\
\sum_{(\boldsymbol{1},\boldsymbol{2})} (q_{B-L})^3&=~-3\;,&
\quad
\sum_{(\boldsymbol{1},\boldsymbol{1}), \,\, Y\neq0} (q_{B-L})^3&=~3\;,
\label{eq:cubic_constraint}
\end{align}
and the quintic,
\begin{align}
\sum_{(\boldsymbol{3},\boldsymbol{2}),(\boldsymbol{\overline{3}},\boldsymbol{2})}
(q_{B-L})^5&=~\frac{1}{81}\;, &
\quad
\sum_{(\boldsymbol{3},\boldsymbol{1}),(\boldsymbol{\overline{3}},\boldsymbol{1})}
(q_{B-L})^5& =~-\frac{2}{81}\;,\nonumber\\
\sum_{(\boldsymbol{1},\boldsymbol{2})}
(q_{B-L})^5& =~-3\;,&
\quad
\sum_{(\boldsymbol{1},\boldsymbol{1}), \,\,Y\neq0}
(q_{B-L})^5& =~3\;,
\label{eq:quintic_constraint}
\end{align}
constraints. As before, the sum over all
$(\boldsymbol{1},\boldsymbol{1})$ representations with non-vanishing
hypercharge reduces to that of the right-handed electrons that carry
$B-L$ charge +1, since the exotic particles come in vector-like
pairs so that their contribution to the sum vanishes.

This leaves us with a set of highly non-linear equations to be
solved. To this end, we used the computer algebra system {\tt
Singular} \cite{GPS05}.  The following cases need to be
distinguished. \begin{inparaenum}[(i)] \item The number of solutions
is finite. \item The solutions are given by continuous parameters,
and the relations intertwining these parameters are linear. \item
The solutions are given by continuous parameters, but this time, the
relations intertwining the parameters are non-linear. $B-L$
directions which lead to irrational charges for the exotics are
discarded in all three cases.
\end{inparaenum}

In the first case, we calculate the spectra for the $B-L$ generators
and check that they are really vector-like. A given model is
particularly interesting if its spectrum contains standard model
singlets with $B-L$ charges $2n/(2m+1)$ for $n,m\in\mathbb{N}$ whose
VEVs break the continuous $\mathrm{U}(1)_{B-L}$ symmetry to a
discrete subgroup which may play the role of a generalized R parity.
In the second case, we use the parametric freedom we have at hand to
assign $B-L$ charges of $2n/(2m+1)$ for $m,n=0,1,2,3$ in all
possible combinations to standard model singlets and thus generate a
list of $B-L$ directions. In the third case, we specialize to a
numerical value and check whether the spectrum satisfies our
criteria. We do not perform a complete search, because these cases
are rare and the analysis more complicated, thus not very rewarding.

\section{Details of Model 1}

The model is defined by the shifts and Wilson lines
\begin{subequations}
\begin{eqnarray}
 V & = &
 \left( \frac{1}{3},-\frac{1}{2},-\frac{1}{2},0,0,0,0,0\right)\,
   \left(\frac{1}{2},-\frac{1}{6},-\frac{1}{2},-\frac{1}{2},-\frac{1}{2},-\frac{1}{2},-\frac{1}{2},\frac{1}{2}\right)\;, \\
 W_2 & = &
 \left( 0,-\frac{1}{2},-\frac{1}{2},-\frac{1}{2},\frac{1}{2},0,0,0\right)\,
   \left(
   4,-3,-\frac{7}{2},-4,-3,-\frac{7}{2},-\frac{9}{2},\frac{7}{2}\right)\;, \\
 W_3 & = &
 \left(-\frac{1}{2},-\frac{1}{2},\frac{1}{6},\frac{1}{6},\frac{1}{6},\frac{1}{6},\frac{1}{6},\frac{1}{6}\right)\,
   \left( \frac{1}{3},0,0,\frac{2}{3},0,\frac{5}{3},-2,0\right)\;.
\end{eqnarray}
\end{subequations}
A possible second order 2 Wilson line is set to zero. The gauge
group after compactification is
\begin{equation}
 G_\mathrm{orbifold}~=~\SU3\times\SU2\times[\SU4\times\SU2']\times\U1^9\;.
\end{equation}

The U(1) generators can be chosen as
\begin{subequations}\label{eq:U1generators1}
\begin{eqnarray}
\mathsf{t}_1 &= & \mathsf{t}_Y~=~
\left(0,0,0,\tfrac{1}{2},\tfrac{1}{2},-\tfrac{1}{3},-\tfrac{1}{3},-\tfrac{1}{3}\right) \, (0,0,0,0,0,0,0,0) \;,\\
\mathsf{t}_2 &= & (1,0,0,0,0,0,0,0) \, (0,0,0,0,0,0,0,0) \;,\\
\mathsf{t}_3 &= & (0,1,0,0,0,0,0,0) \, (0,0,0,0,0,0,0,0) \;,\\
\mathsf{t}_4 &= & (0,0,1,0,0,0,0,0) \, (0,0,0,0,0,0,0,0) \;,\\
\mathsf{t}_5 &= & (0,0,0,1,1,1,1,1) \, (0,0,0,0,0,0,0,0) \;,\\
\mathsf{t}_6 &= & (0,0,0,0,0,0,0,0) \, (0,1,0,0,0,0,0,0) \;,\\
\mathsf{t}_7 &= & (0,0,0,0,0,0,0,0) \, (-1,0,0,1,0,0,0,0) \;,\\
\mathsf{t}_8 &= & (0,0,0,0,0,0,0,0) \, (0,0,0,0,1,0,0,0) \;,\\
\mathsf{t}_9 &= & (0,0,0,0,0,0,0,0) \, (0,0,0,0,0,1,0,0) \;.
\end{eqnarray}
\end{subequations}
The `anomalous' \U1 is generated by
\begin{equation}
 \mathsf{t}_\mathrm{anom}~=~
 \sum\limits_{i=1}^9\alpha_i\,\mathsf{t}_i\;,\quad
 \text{where}\:
 \{\alpha_i\}~=~\left\{0,\tfrac{2}{3},0,-\tfrac{5}{3},\tfrac{1}{3},-\tfrac{1}{3},\tfrac{1}{3},2,\tfrac{1}{3}\right\}
 \;.
\end{equation}
The sum of anomalous charges is
\begin{equation}
 \tr\mathsf{t}_\mathrm{anom}~=~\frac{296}{3}~>~0\;.
\end{equation}

\subsection{Spectrum}

{
\renewcommand{\arraystretch}{1}
\begin{longtable}{|c|c|c|c|c|c|ccc|l|c|c|c|c|c|c|c|c|c|c|c|}
\caption{The spectrum of model 1 in terms of \emph{left-chiral}
states. The \U1 charges refer to the basis of generators
\eqref{eq:U1generators1}. $R_i$ denote $R$-charges.  Note, the
symbol * means the state is not localized in the corresponding
directions. } \label{tab:SpectrumModel1}
\\
\hline
 \!\!\!\!&\!\!\!\! $k$ \!\!\!\!&\!\!\!\! $n_3$ \!\!\!\!&\!\!\!\! $n_2$ \!\!\!\!&\!\!\!\! $n_2'$ \!\!\!\!&\!\!\!\! $q_\gamma$ \!\!\!\!&\!\!\!\!
    $R_1$ \!\!\!\!&\!\!\!\! $R_2$ \!\!\!\!&\!\!\!\! $R_3$ \!\!\!\!&\!\!\!\! $\text{irrep}$
    \!\!\!\!&\!\!\!\! $q_Y$ \!\!\!\!&\!\!\!\! $q_2$ \!\!\!\!&\!\!\!\! $q_3$ \!\!\!\!&\!\!\!\! $q_4$ \!\!\!\!&\!\!\!\! $q_5$ \!\!\!\!&\!\!\!\! $q_6$ \!\!\!\!&\!\!\!\! $q_7$ \!\!\!\!&\!\!\!\! $q_8$ \!\!\!\!&\!\!\!\! $q_9$
    \!\!\!\!&\!\!\!\! $q_{B\!-\!L}$ \!\!\!\!&\!\!\!\! $q_\mathrm{anom}$ \\
\hline
\hline
\endfirsthead
\hline
 \!\!\!\!&\!\!\!\! $k$ \!\!\!\!&\!\!\!\! $n_3$ \!\!\!\!&\!\!\!\! $n_2$ \!\!\!\!&\!\!\!\! $n_2'$ \!\!\!\!&\!\!\!\! $q_\gamma$ \!\!\!\!&\!\!\!\!
    $R_1$ \!\!\!\!&\!\!\!\! $R_2$ \!\!\!\!&\!\!\!\! $R_3$ \!\!\!\!&\!\!\!\! $\text{irrep}$
    \!\!\!\!&\!\!\!\! $q_Y$ \!\!\!\!&\!\!\!\! $q_2$ \!\!\!\!&\!\!\!\! $q_3$ \!\!\!\!&\!\!\!\! $q_4$ \!\!\!\!&\!\!\!\! $q_5$ \!\!\!\!&\!\!\!\! $q_6$ \!\!\!\!&\!\!\!\! $q_7$ \!\!\!\!&\!\!\!\! $q_8$ \!\!\!\!&\!\!\!\! $q_9$
    \!\!\!\!&\!\!\!\! $q_{B\!-\!L}$ \!\!\!\!&\!\!\!\! $q_\mathrm{anom}$ \\
\hline
\hline
\endhead
\hline
\multicolumn{21}{|r|}{{\it continued ...}}\\
\hline
\endfoot

\endlastfoot
$\bar n_{3}$ \!\!\!\!&\!\!\!\! $0$ \!\!\!\!&\!\!\!\! $*$ \!\!\!\!&\!\!\!\! $*$ \!\!\!\!&\!\!\!\! $*$ \!\!\!\!&\!\!\!\! $0$ \!\!\!\!&\!\!\!\! $-1$ \!\!\!\!&\!\!\!\! $0$ \!\!\!\!&\!\!\!\! $0$ \!\!\!\!&\!\!\!\! $\left(\boldsymbol{1},\boldsymbol{1};\boldsymbol{1},\boldsymbol{1}\right)$ \!\!\!\!&\!\!\!\! $0$ \!\!\!\!&\!\!\!\! $\frac{1}{2}$ \!\!\!\!&\!\!\!\! $\frac{1}{2}$ \!\!\!\!&\!\!\!\! $-\frac{1}{2}$ \!\!\!\!&\!\!\!\! $-\frac{5}{2}$ \!\!\!\!&\!\!\!\! $0$ \!\!\!\!&\!\!\!\! $0$ \!\!\!\!&\!\!\!\! $0$ \!\!\!\!&\!\!\!\! $0$ \!\!\!\!&\!\!\!\! $1$ \!\!\!\!&\!\!\!\! $-\frac{1}{3}$\\
$\bar e_{3}$ \!\!\!\!&\!\!\!\! $0$ \!\!\!\!&\!\!\!\! $*$ \!\!\!\!&\!\!\!\! $*$ \!\!\!\!&\!\!\!\! $*$ \!\!\!\!&\!\!\!\! $0$ \!\!\!\!&\!\!\!\! $-1$ \!\!\!\!&\!\!\!\! $0$ \!\!\!\!&\!\!\!\! $0$ \!\!\!\!&\!\!\!\! $\left(\boldsymbol{1},\boldsymbol{1};\boldsymbol{1},\boldsymbol{1}\right)$ \!\!\!\!&\!\!\!\! $1$ \!\!\!\!&\!\!\!\! $\frac{1}{2}$ \!\!\!\!&\!\!\!\! $-\frac{1}{2}$ \!\!\!\!&\!\!\!\! $\frac{1}{2}$ \!\!\!\!&\!\!\!\! $-\frac{1}{2}$ \!\!\!\!&\!\!\!\! $0$ \!\!\!\!&\!\!\!\! $0$ \!\!\!\!&\!\!\!\! $0$ \!\!\!\!&\!\!\!\! $0$ \!\!\!\!&\!\!\!\! $1$ \!\!\!\!&\!\!\!\! $\frac{2}{3}$\\
$\bar f_{1}$ \!\!\!\!&\!\!\!\! $0$ \!\!\!\!&\!\!\!\! $*$ \!\!\!\!&\!\!\!\! $*$ \!\!\!\!&\!\!\!\! $*$ \!\!\!\!&\!\!\!\! $0$ \!\!\!\!&\!\!\!\! $-1$ \!\!\!\!&\!\!\!\! $0$ \!\!\!\!&\!\!\!\! $0$ \!\!\!\!&\!\!\!\! $\left(\boldsymbol{1},\boldsymbol{1};\overline{\boldsymbol{4}},\boldsymbol{1}\right)$ \!\!\!\!&\!\!\!\! $0$ \!\!\!\!&\!\!\!\! $0$ \!\!\!\!&\!\!\!\! $0$ \!\!\!\!&\!\!\!\! $0$ \!\!\!\!&\!\!\!\! $0$ \!\!\!\!&\!\!\!\! $\frac{1}{2}$ \!\!\!\!&\!\!\!\! $-1$ \!\!\!\!&\!\!\!\! $-\frac{1}{2}$ \!\!\!\!&\!\!\!\! $-\frac{1}{2}$ \!\!\!\!&\!\!\!\! $-1$ \!\!\!\!&\!\!\!\! $\frac{5}{3}$\\
\hline
\end{longtable}
}

\subsection{Mass matrices for exotics in Model 1A}

\label{sec:exoticsMasses1}

Here we show that that all exotics can be made massive. The exotic's
mass terms are
\begin{equation}
 x_i\,(\mathcal{M}_{x\bar x})_{ij}\,\bar x_j\;.
\end{equation}
In the following, we list the structure of the corresponding mass
matrices.
\begin{subequations}
\begin{eqnarray}
\mathcal{M}_{\bar\ell\ell} & = & \left(
\begin{array}{cccc}
 0 & 0 & \widetilde{s}^6 & \widetilde{s}^6
\end{array}
\right)\;,\label{eq:Mbarlla}\\
\mathcal{M}_{d\bar d} & = & \left(
\begin{array}{cccc}
 0 & 0 & \widetilde{s}^6 & \widetilde{s}^6
\end{array}
\right)\;,\label{eq:Mdbarda}\\
\mathcal{M}_{mm} & = & \left(
\begin{array}{cccccccc}
 0 & \widetilde{s} & 0 & \widetilde{s}^6 & 0 & 0 & 0 & 0 \\
 \widetilde{s} & 0 & \widetilde{s}^6 & 0 & 0 & 0 & 0 & 0 \\
 0 & \widetilde{s}^6 & 0 & \widetilde{s} & 0 & 0 & 0 & 0 \\
 \widetilde{s}^6 & 0 & \widetilde{s} & 0 & 0 & 0 & 0 & 0 \\
 0 & 0 & 0 & 0 & 0 & 0 & \widetilde{s}^5 & \widetilde{s}^5 \\
 0 & 0 & 0 & 0 & 0 & 0 & \widetilde{s}^5 & \widetilde{s}^5 \\
 0 & 0 & 0 & 0 & \widetilde{s}^5 & \widetilde{s}^5 & 0 & 0 \\
 0 & 0 & 0 & 0 & \widetilde{s}^5 & \widetilde{s}^5 & 0 & 0
\end{array}
\right)\;,\label{eq:Mmma}\\
\mathcal{M}_{s^+s^-} & = & \left(
\begin{array}{cccccccccccccc}
 0 & 0 & 0 & 0 & \widetilde{s}^6 & \widetilde{s}^5 & \widetilde{s}^6 & \widetilde{s}^5 & 0 & 0 & 0 & 0 & 0 & 0 \\
 0 & 0 & 0 & 0 & \widetilde{s}^5 & \widetilde{s} & \widetilde{s}^5 & \widetilde{s}^6 & 0 & 0 & 0 & 0 & 0 & 0 \\
 0 & 0 & 0 & 0 & \widetilde{s}^6 & \widetilde{s}^5 & \widetilde{s}^6 & \widetilde{s}^5 & 0 & 0 & 0 & 0 & 0 & 0 \\
 0 & 0 & 0 & 0 & \widetilde{s}^5 & \widetilde{s}^6 & \widetilde{s}^5 & \widetilde{s} & 0 & 0 & 0 & 0 & 0 & 0 \\
 \widetilde{s}^5 & \widetilde{s}^5 & \widetilde{s}^5 & \widetilde{s}^5 & 0 & 0 & 0 & 0 & 0 & 0 & 0 & 0 & 0 & 0 \\
 0 & \widetilde{s}^6 & 0 & \widetilde{s}^6 & 0 & 0 & 0 & 0 & 0 & 0 & 0 & 0 & 0 & 0 \\
 \widetilde{s}^5 & \widetilde{s}^5 & \widetilde{s}^5 & \widetilde{s}^5 & 0 & 0 & 0 & 0 & 0 & 0 & 0 & 0 & 0 & 0 \\
 0 & \widetilde{s}^6 & 0 & \widetilde{s}^6 & 0 & 0 & 0 & 0 & 0 & 0 & 0 & 0 & 0 & 0 \\
 0 & 0 & 0 & 0 & 0 & 0 & 0 & 0 & \widetilde{s}^2 & \widetilde{s}^2 & \widetilde{s}^6 & 0 & 0 & \widetilde{s}^6 \\
 0 & 0 & 0 & 0 & 0 & 0 & 0 & 0 & \widetilde{s}^2 & \widetilde{s}^2 & \widetilde{s}^6 & 0 & 0 & \widetilde{s}^6 \\
 0 & 0 & 0 & 0 & 0 & 0 & 0 & 0 & \widetilde{s}^6 & \widetilde{s}^6 & 0 & \widetilde{s}^6 & \widetilde{s}^6 & 0 \\
 0 & 0 & 0 & 0 & 0 & 0 & 0 & 0 & 0 & 0 & \widetilde{s}^6 & \widetilde{s}^2 & \widetilde{s}^2 & \widetilde{s}^6 \\
 0 & 0 & 0 & 0 & 0 & 0 & 0 & 0 & 0 & 0 & \widetilde{s}^6 & \widetilde{s}^2 & \widetilde{s}^2 & \widetilde{s}^6 \\
 0 & 0 & 0 & 0 & 0 & 0 & 0 & 0 & \widetilde{s}^6 & \widetilde{s}^6 & 0 & \widetilde{s}^6 & \widetilde{s}^6 & 0
\end{array}
\right)\;,\label{eq:Mspsma}\\
\mathcal{M}_{\delta\bar\delta} & = & \left(
\begin{array}{cccccc}
 0 & 0 & 0 & \widetilde{s}^6 & \widetilde{s}^6 & 0 \\
 0 & \widetilde{s}^5 & \widetilde{s}^5 & 0 & 0 & \widetilde{s}^6 \\
 0 & \widetilde{s}^5 & \widetilde{s}^5 & 0 & 0 & \widetilde{s}^6 \\
 \widetilde{s}^6 & 0 & 0 & \widetilde{s}^6 & \widetilde{s}^6 & 0 \\
 \widetilde{s}^6 & 0 & 0 & \widetilde{s}^6 & \widetilde{s}^6 & 0 \\
 0 & \widetilde{s}^6 & \widetilde{s}^6 & 0 & 0 & 0
\end{array}
\right)\;,\label{eq:Mdeltabardeltaa}\\
\mathcal{M}_{yy} & = & \left(
\begin{array}{cc}
 \widetilde{s} & \widetilde{s}^6 \\
 \widetilde{s}^6 & \widetilde{s}
\end{array}
\right)\;,\label{eq:Myya}\\
\mathcal{M}_{v\bar v} & = & \left(
\begin{array}{cc}
 \widetilde{s} & \widetilde{s}^6 \\
 \widetilde{s}^6 & \widetilde{s}
\end{array}
\right)\;,\label{eq:Mvbarva}\\
\mathcal{M}_{f\bar f} & = & \left(
\begin{array}{cccccc}
 0 & 0 & 0 & 0 & 0 & 0 \\
 0 & \widetilde{s}^5 & \widetilde{s}^5 & \widetilde{s}^5 & \widetilde{s}^5 & 0 \\
 0 & \widetilde{s}^5 & \widetilde{s}^5 & \widetilde{s}^5 & \widetilde{s}^5 & 0 \\
 0 & 0 & 0 & \widetilde{s}^6 & \widetilde{s}^6 & \widetilde{s}^6 \\
 0 & 0 & 0 & \widetilde{s}^6 & \widetilde{s}^6 & \widetilde{s}^6 \\
 0 & 0 & 0 & \widetilde{s}^6 & \widetilde{s}^6 & \widetilde{s}^6
\end{array}
\right)\;,\label{eq:Mfbarfa}\\
\mathcal{M}_{\bar f^+ f^-} & = &
\left(
\begin{array}{cc}
 \widetilde{s}^6 & \widetilde{s}^6 \\
 \widetilde{s}^6 & \widetilde{s}^6
\end{array}
\right)\;.\label{eq:Mbarfpfm}
\end{eqnarray}
\end{subequations}

\subsection{Mass matrices for exotics in Model 1B}

\label{sec:exoticsMasses1b}

Here we show that that all exotics can be made massive. The exotic's
mass terms are
\begin{equation}
 x_i\,(\mathcal{M}_{x\bar x})_{ij}\,\bar x_j\;.
\end{equation}
In the following, we list the structure of the corresponding mass
matrices.
\begin{subequations}
\begin{eqnarray}
\mathcal{M}_{\bar\ell\ell} & = & \left(
\begin{array}{cccc}
 0 & 0 & \widetilde{s}^3 & \widetilde{s}^3
\end{array}
\right)\;,\label{eq:Mbarllb}\\
\mathcal{M}_{d\bar d} & = & \left(
\begin{array}{cccc}
 0 & 0 & \widetilde{s}^3 & \widetilde{s}^3
\end{array}
\right)\;,\label{eq:Mdbardb}\\
\mathcal{M}_{mm} & = & \left(
\begin{array}{cccccccc}
 0 & \widetilde{s} & 0 & \widetilde{s}^6 & 0 & 0 & 0 & 0 \\
 \widetilde{s} & 0 & \widetilde{s}^6 & 0 & 0 & 0 & 0 & 0 \\
 0 & \widetilde{s}^6 & 0 & \widetilde{s} & 0 & 0 & 0 & 0 \\
 \widetilde{s}^6 & 0 & \widetilde{s} & 0 & 0 & 0 & 0 & 0 \\
 0 & 0 & 0 & 0 & 0 & 0 & \widetilde{s}^5 & \widetilde{s}^5 \\
 0 & 0 & 0 & 0 & 0 & 0 & \widetilde{s}^5 & \widetilde{s}^5 \\
 0 & 0 & 0 & 0 & \widetilde{s}^5 & \widetilde{s}^5 & 0 & 0 \\
 0 & 0 & 0 & 0 & \widetilde{s}^5 & \widetilde{s}^5 & 0 & 0
\end{array}
\right)\;,\label{eq:Mmmb}\\
\mathcal{M}_{s^+s^-} & = & \left(
\begin{array}{cccccccccccccc}
 0 & 0 & 0 & 0 & \widetilde{s}^6 & \widetilde{s}^5 & \widetilde{s}^6 & \widetilde{s}^5 & 0 & 0 & 0 & 0 & 0 & 0 \\
 0 & 0 & 0 & 0 & \widetilde{s}^5 & \widetilde{s} & \widetilde{s}^5 & \widetilde{s}^6 & 0 & 0 & 0 & 0 & 0 & 0 \\
 0 & 0 & 0 & 0 & \widetilde{s}^6 & \widetilde{s}^5 & \widetilde{s}^6 & \widetilde{s}^5 & 0 & 0 & 0 & 0 & 0 & 0 \\
 0 & 0 & 0 & 0 & \widetilde{s}^5 & \widetilde{s}^6 & \widetilde{s}^5 & \widetilde{s} & 0 & 0 & 0 & 0 & 0 & 0 \\
 \widetilde{s}^5 & \widetilde{s}^5 & \widetilde{s}^5 & \widetilde{s}^5 & 0 & 0 & 0 & 0 & 0 & 0 & 0 & 0 & 0 & 0 \\
 0 & \widetilde{s}^6 & 0 & \widetilde{s}^6 & 0 & 0 & 0 & 0 & 0 & 0 & 0 & 0 & 0 & 0 \\
 \widetilde{s}^5 & \widetilde{s}^5 & \widetilde{s}^5 & \widetilde{s}^5 & 0 & 0 & 0 & 0 & 0 & 0 & 0 & 0 & 0 & 0 \\
 0 & \widetilde{s}^6 & 0 & \widetilde{s}^6 & 0 & 0 & 0 & 0 & 0 & 0 & 0 & 0 & 0 & 0 \\
 0 & 0 & 0 & 0 & 0 & 0 & 0 & 0 & \widetilde{s} & \widetilde{s}^2 & \widetilde{s}^6 & \widetilde{s}^6 & \widetilde{s}^6 & \widetilde{s}^6 \\
 0 & 0 & 0 & 0 & 0 & 0 & 0 & 0 & \widetilde{s}^2 & \widetilde{s} & \widetilde{s}^6 & \widetilde{s}^6 & \widetilde{s}^6 & \widetilde{s}^6 \\
 0 & 0 & 0 & 0 & 0 & 0 & 0 & 0 & \widetilde{s}^6 & \widetilde{s}^6 & 0 & \widetilde{s}^6 & \widetilde{s}^6 & 0 \\
 0 & 0 & 0 & 0 & 0 & 0 & 0 & 0 & \widetilde{s}^6 & \widetilde{s}^6 & \widetilde{s}^6 & \widetilde{s} & \widetilde{s}^2 & \widetilde{s}^6 \\
 0 & 0 & 0 & 0 & 0 & 0 & 0 & 0 & \widetilde{s}^6 & \widetilde{s}^6 & \widetilde{s}^6 & \widetilde{s}^2 & \widetilde{s} & \widetilde{s}^6 \\
 0 & 0 & 0 & 0 & 0 & 0 & 0 & 0 & \widetilde{s}^6 & \widetilde{s}^6 & 0 & \widetilde{s}^6 & \widetilde{s}^6 & 0
\end{array}
\right)\;,\label{eq:Mspsmb}\\
\mathcal{M}_{\delta\bar\delta} & = & \left(
\begin{array}{cccccc}
 \widetilde{s}^6 & 0 & 0 & \widetilde{s}^3 & \widetilde{s}^3 & 0 \\
 0 & \widetilde{s}^4 & \widetilde{s}^4 & 0 & 0 & \widetilde{s}^3 \\
 0 & \widetilde{s}^4 & \widetilde{s}^4 & 0 & 0 & \widetilde{s}^3 \\
 \widetilde{s}^3 & 0 & 0 & \widetilde{s}^4 & \widetilde{s}^4 & 0 \\
 \widetilde{s}^3 & 0 & 0 & \widetilde{s}^4 & \widetilde{s}^4 & 0 \\
 0 & \widetilde{s}^3 & \widetilde{s}^3 & 0 & 0 & 0
\end{array}
\right)\;,\label{eq:Mdeltabardeltab}\\
\mathcal{M}_{yy} & = & \left(
\begin{array}{cc}
 \widetilde{s} & \widetilde{s}^6 \\
 \widetilde{s}^6 & \widetilde{s}
\end{array}
\right)\;,\label{eq:Myyb}\\
\mathcal{M}_{v\bar v} & = & \left(
\begin{array}{cc}
 \widetilde{s} & \widetilde{s}^6 \\
 \widetilde{s}^6 & \widetilde{s}
\end{array}
\right)\;,\label{eq:Mvbarvb}\\
\mathcal{M}_{f\bar f} & = & \left(
\begin{array}{cccccc}
 \widetilde{s}^5 & 0 & 0 & 0 & 0 & 0 \\
 0 & \widetilde{s}^5 & \widetilde{s}^5 & \widetilde{s}^5 & \widetilde{s}^5 & 0 \\
 0 & \widetilde{s}^5 & \widetilde{s}^5 & \widetilde{s}^5 & \widetilde{s}^5 & 0 \\
 0 & 0 & 0 & \widetilde{s}^3 & \widetilde{s}^3 & \widetilde{s}^5 \\
 0 & 0 & 0 & \widetilde{s}^5 & \widetilde{s}^5 & \widetilde{s}^3 \\
 0 & 0 & 0 & \widetilde{s}^5 & \widetilde{s}^5 & \widetilde{s}^3
\end{array}
\right)\;,\label{eq:Mfbarfb}\\
\mathcal{M}_{\bar f^+ f^-} & = & \left(
\begin{array}{cc}
 \widetilde{s}^6 & \widetilde{s}^6 \\
 \widetilde{s}^6 & \widetilde{s}^6
\end{array}
\right)\;.\label{eq:Mbarfpfm1B}
\end{eqnarray}
\end{subequations}

\section{Details of Model 2}

The model is defined by the shift and Wilson
lines~\cite{Lebedev:2006kn}
\begin{subequations}
\begin{eqnarray}
V & =
&\left(\tfrac{1}{3},\,-\tfrac{1}{2},\,-\tfrac{1}{2},\,0,\,0,\,0,\,0,\,0\right)\left(\tfrac{1}{2},\,-\tfrac{1}{6},\,-\tfrac{1}{2},\,-\tfrac{1}{2},\,-\tfrac{1}{2},\,-\tfrac{1}{2},\,-\tfrac{1}{2},\,\tfrac{1}{2}\right)
\;,\\
W_{2} & =
&\left(\tfrac{1}{4},\,-\tfrac{1}{4},\,-\tfrac{1}{4},-\tfrac{1}{4},\,-\tfrac{1}{4},\,\tfrac{1}{4},\tfrac{1}{4},\,\tfrac{1}{4}\right)\left(1,\,-1,\,-\tfrac{5}{2},\,-\tfrac{3}{2},\,-\tfrac{1}{2},\,-\tfrac{5}{2},\,-\tfrac{3}{2},\,\tfrac{3}{2}\right)
\;,\\
W_{3} & =
&\left(-\tfrac{1}{2},\,-\tfrac{1}{2},\,\tfrac{1}{6},\tfrac{1}{6},\,\tfrac{1}{6},\,\tfrac{1}{6},\tfrac{1}{6},\,\tfrac{1}{6}\right)\left(\tfrac{10}{3},\,0,\,-6,\,-\tfrac{7}{3},\,-\tfrac{4}{3},\,-5,\,-3,\,3\right)\;.
\end{eqnarray}
\end{subequations}

We use the \U1 generators
\begin{subequations}\label{eq:U1generators}
\begin{eqnarray}
\mathsf{t}_1 & = & \mathsf{t}_Y~=~
\left(0,0,0,\tfrac{1}{2},\tfrac{1}{2},-\tfrac{1}{3},-\tfrac{1}{3},-\tfrac{1}{3}\right) \, (0,0,0,0,0,0,0,0) \;,\\
\mathsf{t}_2 & = & (1,0,0,0,0,0,0,0) \, (0,0,0,0,0,0,0,0) \;,\\
\mathsf{t}_3 & = & (0,1,0,0,0,0,0,0) \, (0,0,0,0,0,0,0,0) \;,\\
\mathsf{t}_4 & = & (0,0,1,0,0,0,0,0) \, (0,0,0,0,0,0,0,0) \;,\\
\mathsf{t}_5 & = & (0,0,0,1,1,1,1,1) \, (0,0,0,0,0,0,0,0) \;,\\
\mathsf{t}_6 & = & (0,0,0,0,0,0,0,0) \, (1,0,0,0,0,0,0,0) \;,\\
\mathsf{t}_7 & = & (0,0,0,0,0,0,0,0) \, (0,1,0,0,0,0,0,0) \;,\\
\mathsf{t}_8 & = & (0,0,0,0,0,0,0,0) \, (0,0,0,1,1,0,0,0)\;.
\end{eqnarray}
\end{subequations}
The anomalous U(1) generator can be expressed through a linear
combination of all \U1 generators,
\begin{equation}
 \mathsf{t}_\mathrm{anom}~=~-\sum c_i\,\mathsf{t}_i\;,\quad
 \text{where}\quad
 c_i~=~\left(0,\frac{7}{3},-1,-\frac{5}{3},\frac{1}{3},\frac{2}{3},-\frac{2}{3},-\frac{2}{3}\right)
 \;.
\end{equation}

The sum of anomalous charges is
\begin{equation}
 \tr\mathsf{t}_\mathrm{anom}~=~\frac{416}{3}~>~0\;.
\end{equation}

\subsection{Spectrum}

{
\renewcommand{\arraystretch}{1}
\begin{longtable}{|c|c|c|c|c|c|ccc|l|c|c|c|c|c|c|c|c|c|c|}
\caption{The spectrum of model 2 in terms of \emph{left-chiral}
states. The \U1 charges refer to the basis of generators
\eqref{eq:U1generators}. $R_i$ denote $R$-charges. Note, the symbol
* means the state is not localized in the corresponding directions.
} \label{tab:SpectrumModel2}
\\
\hline
 \!\!\!\!&\!\!\!\! $k$ \!\!\!\!&\!\!\!\! $n_3$ \!\!\!\!&\!\!\!\! $n_2$ \!\!\!\!&\!\!\!\! $n_2'$ \!\!\!\!&\!\!\!\! $q_\gamma$ \!\!\!\!&\!\!\!\!
        $R_1$ \!\!\!\!&\!\!\!\! $R_2$ \!\!\!\!&\!\!\!\! $R_3$ \!\!\!\!&\!\!\!\! $\text{irrep}$
        \!\!\!\!&\!\!\!\! $q_Y$ \!\!\!\!&\!\!\!\! $q_2$ \!\!\!\!&\!\!\!\! $q_3$ \!\!\!\!&\!\!\!\! $q_4$ \!\!\!\!&\!\!\!\! $q_5$ \!\!\!\!&\!\!\!\! $q_6$ \!\!\!\!&\!\!\!\! $q_7$ \!\!\!\!&\!\!\!\! $q_8$ \!\!\!\!&\!\!\!\! $q_{B\!-\!L}$ \!\!\!\!&\!\!\!\! $q_\mathrm{anom}$\\
\hline
\hline
\endfirsthead
\hline
 \!\!\!\!&\!\!\!\! $k$ \!\!\!\!&\!\!\!\! $n_3$ \!\!\!\!&\!\!\!\! $n_2$ \!\!\!\!&\!\!\!\! $n_2'$ \!\!\!\!&\!\!\!\! $q_\gamma$ \!\!\!\!&\!\!\!\!
        $R_1$ \!\!\!\!&\!\!\!\! $R_2$ \!\!\!\!&\!\!\!\! $R_3$ \!\!\!\!&\!\!\!\! $\text{irrep}$
        \!\!\!\!&\!\!\!\! $q_Y$ \!\!\!\!&\!\!\!\! $q_2$ \!\!\!\!&\!\!\!\! $q_3$ \!\!\!\!&\!\!\!\! $q_4$ \!\!\!\!&\!\!\!\! $q_5$ \!\!\!\!&\!\!\!\! $q_6$ \!\!\!\!&\!\!\!\! $q_7$ \!\!\!\!&\!\!\!\! $q_8$ \!\!\!\!&\!\!\!\! $q_{B\!-\!L}$ \!\!\!\!&\!\!\!\! $q_\mathrm{anom}$\\
\hline
\hline
\endhead
\hline
\multicolumn{20}{|r|}{{\it continued ...}}\\
\hline
\endfoot

\endlastfoot
 $q_{3}$ \!\!\!\!&\!\!\!\! $0$ \!\!\!\!&\!\!\!\! $*$ \!\!\!\!&\!\!\!\! $*$ \!\!\!\!&\!\!\!\! $*$ \!\!\!\!&\!\!\!\! $0$ \!\!\!\!&\!\!\!\! $-1$ \!\!\!\!&\!\!\!\! $0$ \!\!\!\!&\!\!\!\! $0$ \!\!\!\!&\!\!\!\!
 $\left(\boldsymbol{3},\boldsymbol{2};\boldsymbol{1},\boldsymbol{1}\right)$ \!\!\!\!&\!\!\!\! $\frac{1}{6}$ \!\!\!\!&\!\!\!\!
 $\frac{1}{2}$ \!\!\!\!&\!\!\!\! $-\frac{1}{2}$ \!\!\!\!&\!\!\!\! $\frac{1}{2}$ \!\!\!\!&\!\!\!\! $-\frac{1}{2}$ \!\!\!\!&\!\!\!\! $0$ \!\!\!\!&\!\!\!\! $0$ \!\!\!\!&\!\!\!\! $0$ \!\!\!\!&\!\!\!\! $\frac{1}{3}$ \!\!\!\!&\!\!\!\! $-\frac{2}{3}$\\
 $\chi_{1}$ \!\!\!\!&\!\!\!\! $0$ \!\!\!\!&\!\!\!\! $*$ \!\!\!\!&\!\!\!\! $*$ \!\!\!\!&\!\!\!\! $*$ \!\!\!\!&\!\!\!\! $0$ \!\!\!\!&\!\!\!\! $-1$ \!\!\!\!&\!\!\!\! $0$ \!\!\!\!&\!\!\!\! $0$ \!\!\!\!&\!\!\!\! $\left(\boldsymbol{1},\boldsymbol{1};\boldsymbol{1},\boldsymbol{1}\right)$ \!\!\!\!&\!\!\!\! $0$ \!\!\!\!&\!\!\!\! $-1$ \!\!\!\!&\!\!\!\! $-1$ \!\!\!\!&\!\!\!\! $0$ \!\!\!\!&\!\!\!\! $0$ \!\!\!\!&\!\!\!\! $0$ \!\!\!\!&\!\!\!\! $0$ \!\!\!\!&\!\!\!\! $0$ \!\!\!\!&\!\!\!\! $-2$ \!\!\!\!&\!\!\!\! $\frac{4}{3}$\\
 $\chi_{2}$ \!\!\!\!&\!\!\!\! $0$ \!\!\!\!&\!\!\!\! $*$ \!\!\!\!&\!\!\!\! $*$ \!\!\!\!&\!\!\!\! $*$ \!\!\!\!&\!\!\!\! $0$ \!\!\!\!&\!\!\!\! $-1$ \!\!\!\!&\!\!\!\! $0$ \!\!\!\!&\!\!\!\! $0$ \!\!\!\!&\!\!\!\! $\left(\boldsymbol{1},\boldsymbol{1};\boldsymbol{1},\boldsymbol{1}\right)$ \!\!\!\!&\!\!\!\! $0$ \!\!\!\!&\!\!\!\! $\frac{1}{2}$ \!\!\!\!&\!\!\!\! $\frac{1}{2}$ \!\!\!\!&\!\!\!\! $-\frac{1}{2}$ \!\!\!\!&\!\!\!\! $-\frac{5}{2}$ \!\!\!\!&\!\!\!\! $0$ \!\!\!\!&\!\!\!\! $0$ \!\!\!\!&\!\!\!\! $0$\!\!\!\!&\!\!\!\! $2$ \!\!\!\!&\!\!\!\! $-\frac{2}{3}$\\
\hline
\end{longtable}
}

\subsection{Mass matrices} \label{sec:exoticsMasses2}

Here we show that that all exotics can be made massive. The exotic's
mass terms are
\begin{equation}
 x_i\,(\mathcal{M}_{x\bar x})_{ij}\,\bar x_j\;.
\end{equation}
In the following, we list the structure of the corresponding mass
matrices.
\begin{subequations}
\begin{eqnarray}
\mathcal{M}_{\bar\ell\ell} & = & \left(
\begin{array}{cccc}
 \widetilde{s}^2 & \widetilde{s}^2 & \widetilde{s}^3 & \widetilde{s}^3
\end{array}
\right)\;,\label{eq:Mbarll}\\
\mathcal{M}_{d\bar d} & = & \left(
\begin{array}{cccc}
 \widetilde{s}^6 & \widetilde{s}^6 & \widetilde{s}^3 & \widetilde{s}^3 \end{array} \right)\;,\label{eq:Mdbard}\\
\mathcal{M}_{mm} & = & \left(
\begin{array}{cccc}
 0 & 0 & \widetilde{s}^6 & \widetilde{s}^6 \\
 0 & 0 & \widetilde{s}^6 & \widetilde{s}^6 \\
 \widetilde{s}^6 & \widetilde{s}^6 & 0 & \widetilde{s}^6 \\
 \widetilde{s}^6 & \widetilde{s}^6 & \widetilde{s}^6 & 0
\end{array}
\right)\;,\label{eq:Mmm}\\
\mathcal{M}_{\delta\bar\delta} & = & \left(
\begin{array}{ccc}
 \widetilde{s}^3 & \widetilde{s}^3 & \widetilde{s}^3 \\
 \widetilde{s}^3 & \widetilde{s}^3 & \widetilde{s}^3 \\
 0 & \widetilde{s}^3 & \widetilde{s}^3
\end{array}
\right)\;,\label{eq:Mdeltabardelta}\\
\mathcal{M}_{yy} & = & \left(
\begin{array}{cc}
 \widetilde{s}^1 & \widetilde{s}^5 \\
 \widetilde{s}^5 & \widetilde{s}^1
\end{array}
\right)\;,\label{eq:Myy}\\
\mathcal{M}_{v\bar v} & = & \left(
\begin{array}{cccc}
 \widetilde{s} & \widetilde{s}^5 & 0 & 0\\
 \widetilde{s}^5 & \widetilde{s} & 0 & 0\\
 0 & 0 & \widetilde{s}^5 & \widetilde{s}^5\\
 0 & 0 & \widetilde{s}^5 & \widetilde{s}^5
\end{array}
\right)\;,\label{eq:Mvbarv}\\
\mathcal{M}_{x^+x^-} & = & \left(
\begin{array}{cc}
 \widetilde{s}^5 & \widetilde{s}^5 \\
 \widetilde{s}^5 & \widetilde{s}^5
\end{array}
\right)\;,\label{eq:Mxpxm}\\
\mathcal{M}_{f\bar f} & = & \left(
\begin{array}{cc}
 0 & \widetilde{s}^3 \\
 0 & \widetilde{s}^3
\end{array}
\right)\;,\label{eq:Mfbarf}\\
\mathcal{M}_{ww} & = & \left(
\begin{array}{ccccc}
 \widetilde{s} & \widetilde{s}^5 & 0 & \widetilde{s}^5 & \widetilde{s}^5\\
 \widetilde{s}^5 & \widetilde{s} & 0 & \widetilde{s}^5 & \widetilde{s}^5\\
 0 & 0 & 0 & \widetilde{s}^3 & \widetilde{s}^3\\
 \widetilde{s}^5 & \widetilde{s}^5 & \widetilde{s}^3 & \widetilde{s}^6 & \widetilde{s}^6\\
 \widetilde{s}^5 & \widetilde{s}^5 & \widetilde{s}^3 & \widetilde{s}^6 & \widetilde{s}^6
\end{array}
\right)\;,\label{eq:Mww}\\
\mathcal{M}_{s^+s^-} & = & \left(
\begin{array}{cccccccccccccccccccc}
 0 \!\!&\!\! 0 \!\!&\!\! 0 \!\!&\!\! 0 \!\!&\!\! 0 \!\!&\!\! 0 \!\!&\!\! 0 \!\!&\!\! 0 \!\!&\!\! 0 \!\!&\!\! 0 \!\!&\!\! 0 \!\!&\!\! 0 \!\!&\!\! 0 \!\!&\!\! \widetilde{s}^6 \!\!&\!\! \widetilde{s}^6 \!\!&\!\! \widetilde{s}^6 \!\!&\!\! 0 \!\!&\!\! \widetilde{s}^6 \!\!&\!\! \widetilde{s}^6 \!\!&\!\! \widetilde{s}^6 \\
 \widetilde{s}^6 \!\!&\!\! 0 \!\!&\!\! \widetilde{s}^6 \!\!&\!\! \widetilde{s}^6 \!\!&\!\! 0 \!\!&\!\! \widetilde{s}^6 \!\!&\!\! 0 \!\!&\!\! 0 \!\!&\!\! 0 \!\!&\!\! 0 \!\!&\!\! 0 \!\!&\!\! \widetilde{s}^6 \!\!&\!\! \widetilde{s}^6 \!\!&\!\! 0 \!\!&\!\! 0 \!\!&\!\! 0 \!\!&\!\! \widetilde{s}^6 \!\!&\!\! 0 \!\!&\!\! 0 \!\!&\!\! 0 \\
 \widetilde{s}^6 \!\!&\!\! 0 \!\!&\!\! 0 \!\!&\!\! \widetilde{s}^6 \!\!&\!\! 0 \!\!&\!\! 0 \!\!&\!\! 0 \!\!&\!\! 0 \!\!&\!\! 0 \!\!&\!\! 0 \!\!&\!\! \widetilde{s}^6 \!\!&\!\! 0 \!\!&\!\! \widetilde{s}^6 \!\!&\!\! 0 \!\!&\!\! 0 \!\!&\!\! 0 \!\!&\!\! \widetilde{s}^6 \!\!&\!\! 0 \!\!&\!\! 0 \!\!&\!\! 0 \\
 0 \!\!&\!\! 0 \!\!&\!\! 0 \!\!&\!\! 0 \!\!&\!\! 0 \!\!&\!\! 0 \!\!&\!\! 0 \!\!&\!\! 0 \!\!&\!\! 0 \!\!&\!\! 0 \!\!&\!\! 0 \!\!&\!\! 0 \!\!&\!\! 0 \!\!&\!\! \widetilde{s}^6 \!\!&\!\! \widetilde{s}^6 \!\!&\!\! \widetilde{s}^6 \!\!&\!\! 0 \!\!&\!\! \widetilde{s}^6 \!\!&\!\! \widetilde{s}^6 \!\!&\!\! \widetilde{s}^6 \\
 \widetilde{s}^6 \!\!&\!\! 0 \!\!&\!\! \widetilde{s}^6 \!\!&\!\! \widetilde{s}^6 \!\!&\!\! 0 \!\!&\!\! \widetilde{s}^6 \!\!&\!\! 0 \!\!&\!\! 0 \!\!&\!\! \widetilde{s}^6 \!\!&\!\! 0 \!\!&\!\! 0 \!\!&\!\! 0 \!\!&\!\! \widetilde{s}^6 \!\!&\!\! 0 \!\!&\!\! 0 \!\!&\!\! 0 \!\!&\!\! \widetilde{s}^6 \!\!&\!\! 0 \!\!&\!\! 0 \!\!&\!\! 0 \\
 \widetilde{s}^6 \!\!&\!\! 0 \!\!&\!\! 0 \!\!&\!\! \widetilde{s}^6 \!\!&\!\! 0 \!\!&\!\! 0 \!\!&\!\! 0 \!\!&\!\! \widetilde{s}^6 \!\!&\!\! 0 \!\!&\!\! 0 \!\!&\!\! 0 \!\!&\!\! 0 \!\!&\!\! \widetilde{s}^6 \!\!&\!\! 0 \!\!&\!\! 0 \!\!&\!\! 0 \!\!&\!\! \widetilde{s}^6 \!\!&\!\! 0 \!\!&\!\! 0 \!\!&\!\! 0 \\
\end{array}
\right)\;.\label{eq:Mspsm}
\end{eqnarray}
\end{subequations}


\providecommand{\bysame}{\leavevmode\hbox to3em{\hrulefill}\thinspace}

\end{document}